\documentclass[aps,prx,reprint,superscriptaddress,longbibliography, floatfix]{revtex4-2}

\usepackage{amsmath,amssymb,amsfonts}
\usepackage{graphicx}
\usepackage{textcomp}
\usepackage{xcolor}
\usepackage{physics}
\usepackage{bm}
\usepackage{enumitem}
\usepackage[caption=false]{subfig}
\usepackage{tikz}
\usetikzlibrary{positioning,backgrounds,decorations.pathreplacing,calc,arrows.meta}
\usepackage{quantikz}

\DeclareRobustCommand{\erase}{\bgroup\markoverwith{\textcolor{red}{\rule[.5ex]{2pt}{0.4pt}}}\ULon}

\begin{document}

\title{Geometric Quantum Physics Informed Neural Network}
\author{Wai-Hong Tam}
\email{w.tam@j-ij.com}
\affiliation{\textit{JIJ Inc.}, 3-3-6 Shibaura, Minato-ku, Tokyo, 108-0023, Japan}
\author{Reza Safari}
\email{reza.safari@dlr.de}
\affiliation{German Aerospace Center (DLR), Institute of Software Methods for Product Virtualization, 01187 Dresden, Germany}
\author{Hiromichi Matsuyama}
\email{h.matsuyama@j-ij.com}
\affiliation{\textit{JIJ Inc.}, 3-3-6 Shibaura, Minato-ku, Tokyo, 108-0023, Japan}

\begin{abstract}
Quantum physics-informed neural networks (QPINNs) have recently emerged as a promising framework for the solution of partial differential equations (PDEs), with several studies reporting improved convergence and accuracy relative to classical physics-informed neural networks (PINNs) at reduced training cost. Motivated by these advances, we introduce geometric quantum physics-informed neural networks (GQPINNs), a symmetry-aware extension of QPINNs in which the geometric structure of the underlying PDE is incorporated directly into the quantum-circuit ansatz. Building on the framework of geometric quantum machine learning, we construct parametrized circuits that encode finite-group and compact Lie-group symmetries as inductive biases through problem-specific equivariant generator sets . Using a twirling-based construction, we derive symmetry-preserving gates that ensure that the model predictions respect the symmetries of the governing equation whenever the boundary and initial data are symmetry compatible. We benchmark GQPINNs against standard QPINNs and symmetry-adapted classical PINN baselines under matched training protocols across a representative set of linear and nonlinear PDEs. Across these benchmarks, GQPINNs achieve improved solution accuracy, as quantified by lower mean absolute error, while requiring substantially fewer trainable parameters. These results identify symmetry-aware quantum-circuit design as an effective route toward improved efficiency and generalization in quantum PDE solvers and provide a systematic framework for incorporating geometric inductive biases into quantum-enhanced scientific machine learning.
\end{abstract}

\maketitle

\section{Introduction}
Quantum computing is being actively investigated as a computational framework for problems in quantum many-body physics, optimization, chemistry, and scientific computing~\cite{Ayral2023ManyBody,Blekos2024QAOAReview,Bauer2020QuantumChemistry,Montanaro2016Overview}. Within scientific computing, the solution of differential equations has emerged as a particularly important target because of its foundational role across a broad range of disciplines. Representative applications include computational fluid dynamics (CFD), engineering design, and geophysical and climate modeling. On classical hardware, such problems are typically addressed using mesh-based discretization techniques, most notably finite-difference, finite-element, and finite-volume methods~\cite{Grossmann2007CFD,Fallah2000CFD}, which reduce the governing equations to large-scale algebraic systems. Although these approaches are mature and highly successful, their computational cost can become prohibitive in multiscale, high-dimensional, or long-time settings. This has motivated sustained interest in quantum algorithms for differential equations in both fault-tolerant and near-term regimes.

In the fault-tolerant setting, several quantum algorithms are known to offer polynomial speedups and, under favorable structural assumptions, potentially superpolynomial advantages for certain classes of differential equations. For linear differential equations, a major line of work builds on quantum linear-systems algorithms (QLSAs), beginning with the Harrow-Hassidim-Lloyd (HHL) algorithm, which uses phase estimation to prepare a quantum state whose amplitudes encode the solution of a sparse linear system~\cite{Harrow_2009}. This framework has subsequently been extended to high-precision solvers for linear PDEs and to time-dependent linear dynamics through linear system~\cite{Childs2021PDE,BerryCosta2024Dyson,Ambainis2012QLS,Childs2017QLS,Montanaro2016QLS,Wang2024QLS}, spectral~\cite{Childs2020QSM}, and quantum-simulation-based formulations~\cite{Childs2021PDE,BerryCosta2024Dyson,Jin2024Schrodingerization,Berry2014HS,Berry2015HS,Costa2019HS}. For nonlinear differential equations, proposed approaches include linearization-based strategies such as Carleman-type embeddings, which map nonlinear dynamics to enlarged linear systems~\cite{Liu2021Nonlinear}, quantum amplitude-estimation-based methods~\cite{Oz2022QAE,Gaitan2020QAE}, and quantum lattice Boltzmann formulations~\cite{Itani2023QLBM,Budinski2022QLBM}. At present, however, these algorithms generally require large-scale fault-tolerant quantum computation (FTQC)~\cite{Preskill2018quantumcomputingin,RevModPhys.94.015004}.

By contrast, currently available quantum processors operate in the noisy intermediate-scale quantum (NISQ) regime, where limited qubit counts, decoherence, and gate imperfections preclude large-scale fault-tolerant computation~\cite{Preskill2018quantumcomputingin,RevModPhys.94.015004}. This setting has motivated the development of hybrid quantum-classical algorithms based on parametrized quantum circuits~\cite{McClean_2016}, which remain among the most widely studied candidates for near-term applications. In the context of differential equations, several such methods have been proposed for both linear~\cite{Demirdjian2022VQAlinear,BravoPrieto2023VQAlinear} and nonlinear PDEs~\cite{Pool2024VQAnonlinear,Lubasch2020VQAnonlinear}, as well as formulations based on variational quantum lattice Boltzmann methods~\cite{Syamlal2024VQLB} and quantum reservoir computing~\cite{Pfeffer2022QRC,Pfeffer2023QRC}.

Within the NISQ setting, physics-informed quantum circuits, also referred to as quantum physics-informed neural networks (QPINNs)~\cite{kyriienko2021PIQC, Siegl2025PIQC, klement2026PIQC}, have recently been investigated as quantum counterparts of classical physics-informed neural networks (PINNs). In the PINN framework, the governing equations, together with boundary and initial conditions, are incorporated directly into the training objective~\cite{Raissi2019PINN,Cai2021PINN,Wassing2024PINN,Karniadakis2021PINN}. Because PINNs parametrize continuous solution fields directly, they provide a conceptually mesh-free approach that is particularly attractive for inverse problems and irregular domains. QPINNs extend this paradigm to variational quantum machine learning, in which a parametrized quantum circuit serves as an ansatz for the solution field and its parameters are optimized by a classical routine.

A central challenge in QPINNs is that increasingly complex PDEs typically require more expressive variational quantum circuits and, consequently, a larger number of trainable parameters~\cite{Siegl2025PIQC}. Increasing the parameter count, however, generally renders such models more difficult to optimize. Moreover, variational quantum circuits may suffer from barren plateaus~\cite{McClean2018BarrenPlateaus,Qi2023BarrenPlateaus,Larocca2025BarrenPlateaus}. These observations underscore the importance of designing circuit architectures that reflect the underlying structure of the problem. In classical machine learning, geometric deep learning provides a framework for incorporating inductive biases---such as symmetry, invariance, and equivariance---directly into model design, often improving data efficiency, optimization, and generalization~\cite{Bronstein2017GML,bronstein2021GML,Cohen2016GML}. Closely related ideas have recently been extended to quantum models under the framework of geometric quantum machine learning~\cite{ragone2022GQML,larocca2023groupequivariant,Nguyen2024equivariantQC,Meyer2023QML}. Since symmetry often plays a fundamental role in the structure and analysis of PDEs, these developments suggest that QPINNs should benefit from circuit architectures in which symmetry constraints are built in by construction, rather than learned implicitly during training.

In this work, we develop a geometric QPINN (GQPINN) framework for PDEs with spatial symmetries. Our approach embeds the target solution symmetry directly into the quantum-circuit architecture. We apply this framework to several linear and nonlinear PDEs and benchmark its performance against a baseline QPINN ansatz introduced in prior work~\cite{Siegl2025PIQC}. Across all benchmarks considered, the GQPINN ansatz consistently outperforms the baseline QPINN. For the two-dimensional Poisson problem, GQPINN achieves up to two orders of magnitude lower mean absolute error while using fewer trainable parameters. For the two-dimensional diffusion equation, GQPINN continues to improve as the number of parameters increases, whereas the baseline QPINN largely saturates. For the one-dimensional acoustic wave equation, the GQPINN reduces the mean absolute error by more than one order of magnitude relative to QPINN at moderate parameter counts. For the one-dimensional viscous Burgers equation, GQPINN again performs consistently better than QPINN, although with a more modest margin. Taken together, these results show that embedding symmetry into the circuit architecture can substantially improve the performance of QPINNs on representative PDE problems.

The remainder of this paper is organized as follows. Section~\ref{sec:background} reviews the theoretical background on quantum machine learning, geometric quantum machine learning and QPINNs. Section~\ref{sec:group_rep} introduces the symmetry-embedded circuit construction for $K_4$ and $SO(2)$ (rotational) symmetries and discusses its extension to the two-dimensional diffusion equation. Section~\ref{sec:experimental_result} presents the numerical benchmarks for the Poisson and diffusion problems, including comparisons with the baseline QPINN ansatz. We also discuss equivariant models in Sec.~\ref{app:Z_2_case}. 
In Sec.~\ref{sec:expressibility}, we theoretically analyze the expressibility of the considered QPINN and GQPINN ansatzes by computing the KL-divergence between their fidelity distributions and the Haar-random distribution.
Section~\ref{sec:summary} summarizes the main results and outlines directions for future work.

\section{Background}
\label{sec:background}

\subsection{Quantum Machine Learning}
The goal of a machine-learning model is to learn the parameters \( \bm{\theta} \) of a model \( f_{\bm{\theta}} \) by minimizing a loss function on a training dataset so as to achieve accurate predictions on unseen data.
Quantum machine learning (QML) comprises machine learning methods implemented with quantum circuits, in which data are encoded into quantum states, processed by parametrized quantum gates, and mapped to predictions through measurement outcomes.
As in classical machine learning, a variety of QML models have been proposed for tasks such as classification, time-series forecasting, and reinforcement learning.
Although QML can treat both quantum and classical inputs, here we focus on QML with classical inputs.
This setting is natural for QPINNs, where the inputs are the space-time coordinates \(\bm{z}_i = (\bm{x}_i,t_i)\).

In QML, we consider a quantum model \( f_{\bm{\theta}} : \mathcal{Z} \to \mathcal{Y} \), where \(\mathcal{Z}\) and \(\mathcal{Y}\) denote the input and output domains, respectively.
Here we consider variational quantum machine-learning models,
\begin{equation}
    y_i = f_{\bm{\theta}}(\bm{z}_i) = \expval{O}{\psi_{\bm{\theta}}(\bm{z}_{i})},
    \label{eq:qml_exp}
\end{equation}
where \(O\) is a chosen observable, and \(\bm{z}_i \in \mathcal{Z}\) and \(y_i \in \mathcal{Y}\) are the \(i\)th input and corresponding prediction, respectively.
The model output is a classical value obtained by measuring the observable \(O\) on the prepared quantum state.
In practice, the expectation value is estimated from a finite number of measurement shots by averaging the measurement outcomes, which introduces statistical noise.
The parametrized quantum state \(\ket{\psi(\bm{\theta}, \bm{z})}\) is prepared from an initial state \(\ket{\psi_0}\) by a parametrized circuit \(U(\bm{\theta}, \bm{z})\).

The design of a parametrized circuit is central to variational quantum algorithms (VQAs).
In this study, we consider the data re-uploading method~\cite{Schuld2021reupload,perez-salinas2020reupload,jerbi2023reupload} for the parametrized quantum circuit.
In this approach, the input data and trainable parameters are encoded into separate circuit blocks: a data-encoding block \(U(\bm{z})\) and trainable blocks \(W_i(\bm{\theta})\).
The blocks \(U(\bm{z})\) and \(W_i(\bm{\theta})\) are repeated \(p\) and \(p+1\) times, respectively, such that,
\begin{equation}
\begin{split}
    \ket{\psi(\boldsymbol{\theta}, \boldsymbol{z})} &= U(\bm{\theta}, \bm{z})\ket{\psi_0} \\
    &=  W_{p+1}(\boldsymbol{\theta}) U(\boldsymbol{z}) \cdots W_2(\boldsymbol{\theta}) U(\boldsymbol{z}) W_1(\bm{\theta})\ket{\psi_0}.
\end{split}
\label{eq:ansatzstate}
\end{equation}
We train the parametrized quantum circuit by minimizing the loss function \(L(\bm{\theta})\).
After obtaining the optimal parameters \(\bm{\theta}^*\), we use $f_{\bm{\theta}^*}(\bm{z}) $ as the model prediction for input \(\bm{z}\).
Although several constructions of the data-encoding unitary \(U(\bm{z})\) are possible.
The design of trainable blocks that exploit symmetry is discussed in the next subsection.

\subsection{Geometric Quantum Machine Learning \label{sec:gqml}}
The architecture of a parametrized quantum circuit is a central factor in determining the expressibility, trainability, and generalization performance of a quantum machine learning model.
Geometric quantum machine learning (GQML) incorporates geometric structure (in particular, symmetries) as inductive biases in quantum-circuit models.
In our setting, we embed these symmetries through the design of trainable block $W_i(\bm{\theta})$.
Circuit constructions that respect symmetries are discussed in Ref.~\cite{Meyer2023QML}, and we follow that framework here.
We assume that our prediction $y$ is invariant under a symmetry group $\mathcal{G}$ that acts on the input space $\mathcal{Z}$ (e.g. rotations).
Let $V: \mathcal{G} \to \text{Aut}(\mathcal{Z})$ denotes the group action on $\mathcal{Z}$, with $V_g \in \text{Aut}(\mathcal{Z})$ for $g \in \mathcal{G}$.
The invariance condition can be written as:
\begin{equation}
f_{\bm{\theta}}(V_g[\boldsymbol{z}]) = f_{\bm{\theta}}(\boldsymbol{z}) \quad \forall \boldsymbol{z} \in \mathcal{Z}, g \in \mathcal{G}.
\end{equation}
Equivalently, Eq.~\eqref{eq:qml_exp} implies:
\begin{equation}
\begin{split}
    &\expval{O}{\psi_{\bm{\theta}}(V_g[\bm{z}])}\\
    &= \expval{O}{\psi_{\bm{\theta}}( \bm{z})} \quad
    \forall \bm{z}\in \mathcal{Z}, g \in \mathcal{G}.
\end{split}
\label{eq:invariance_eq}
\end{equation}
We refer to a model that satisfies Eq.~\eqref{eq:invariance_eq} as a $\mathcal{G}$-invariant model. 
Note that we can also define a $\mathcal{G}$-equivariant model by introducing an action $K:\mathcal{G} \to \text{Aut}(\mathcal{Y})$ on the output space and requiring
\begin{equation}
f_{\bm{\theta}}(V_g[\boldsymbol{z}]) = K_g\,f_{\bm{\theta}}(\boldsymbol{z})
\quad \forall \boldsymbol{z}\in\mathcal{Z},\ g\in\mathcal{G}.
\label{eq:equivariance_eq}
\end{equation}
In terms of Eq.~\eqref{eq:qml_exp}, this becomes \begin{equation}
\begin{split}
    &\expval{O}{\psi_{\bm{\theta}}(V_g[\bm{z}])}\\
    &= K_g \expval{O}{\psi_{\bm{\theta}}( \bm{z})} \quad
    \forall \bm{z}\in \mathcal{Z}, g \in \mathcal{G}.
\end{split}
\label{eq:equivariance_exp}
\end{equation}

All components of a $\mathcal{G}$-invariant model should respect the input symmetry $V_g$.
We call the data-encoding unitary $U(\bm{z})$ $\mathcal{G}$-equivariant if it satisfies,
\begin{equation}
    U(V_g[\bm{z}]) = U_g U(\bm{z}) U_g^\dagger.
\label{eq:eqvembed}
\end{equation}
where $U_g$ is a unitary representation of $g\in\mathcal{G}$ on the Hilbert space induced by the data encoding.

In this study, we consider the data re-uploading ansatz $U(\bm{\theta},\bm{z})$ defined in Eq.~\eqref{eq:ansatzstate}.
Using the $\mathcal{G}$-equivariant data encoding in Eq.~\eqref{eq:eqvembed}, we assume that each trainable block commutes with \(U_g\) for all $g \in \mathcal{G}$,
\begin{equation}
    [W_i(\bm{\theta}), U_g] = 0 \quad \forall g \in \mathcal{G}.
\label{eq:commugen}
\end{equation}
Then the full circuit transforms covariantly as:
\begin{equation}
    U(\bm{\theta}, V_g[\bm{z}]) = U_g\, U(\bm{\theta},\bm{z})\, U_g^\dagger,
\end{equation}
hence, the corresponding state satisfies:
\begin{equation}
\ket{\psi(\bm{\theta}, V_g[\bm{z}])} = U_g U(\bm{\theta}, \bm{z})U_g^\dagger \ket{\psi_0}.
\end{equation}
The expectation value becomes,
\begin{equation}
\begin{split}
    &\expval{O}{\psi(\bm{\theta}, V_g[\bm{z}])} \\
    =& \bra{ \psi_0 } U_g U^\dagger(\bm{\theta}, \bm{z})U_g^\dagger OU_g U(\bm{\theta}, \bm{z})U_g^\dagger \ket{\psi_0}
\end{split}
\end{equation}
If we choose an invariant initial state and an invariant observable such that:
\begin{equation}
    U_g \ket{\psi_0} = e^{i\phi(g)}\ket{\psi_0},\ U_g^\dagger O U_g = O,
\label{eq:state-invariance}
\end{equation}
then the model satisfies Eq.~\eqref{eq:invariance_eq}, and becomes $\mathcal{G}$-invariant.

To construct a $\mathcal{G}$-invariant model, we design the trainable block $W_i(\bm{\theta})$ to satisfy Eq.~\eqref{eq:commugen}.
We assume that the generator set  $\mathcal{W}$ used in each trainable block consists of exponentials of Hermitian generators,
\begin{equation}
    \mathcal{W} = \{e^{-i\theta H}\mid H \in \mathcal{S}\}
\end{equation}
where $\mathcal{S}$ is a given set of Hermitian generators and $\theta$ is a trainable parameter.
This exponential form is commonly used in variational quantum circuits.
A sufficient way to enforce Eq.~\eqref{eq:commugen} is to choose generators satisfying:
\begin{equation}
    [H, U_g] =0\quad \forall g \in \mathcal{G}.
    \label{eq:commu_generator}
\end{equation}
Constructing a set of generators that satisfies Eq.~\eqref{eq:commu_generator} allows us to build $\mathcal{G}$-invariant trainable block.
The twirling formula~\cite{Helsen2021twirling},
\begin{equation}
T_{\mathcal{G}}(H) = \frac{1}{|\mathcal{G}|} \sum_{g \in \mathcal{G}} U_g H U_g^\dagger,
\label{eq:twirling}
\end{equation}
maps $H$ to an operator that commutes with $U_g$ for all $g\in \mathcal{G}$, and therefore,
\begin{equation}
    [T_{\mathcal{G}}(H), U_g]=0 \quad \forall g \in \mathcal{G}.
\end{equation}
Accordingly, for continuous (compact) groups the twirling formula is obtained by replacing the sum with an integral over the normalized Haar measure $\mu$,
\begin{equation}
    T_{\mathcal{G}}(H) = \int d\mu(g) U_g H U_g^\dagger.
\label{eq:twirling_cont}
\end{equation}
This construction induces a commuting generator set,
\begin{equation}
    \mathcal{T}_{\mathcal{G}}(\mathcal{S}) = \{ T_{\mathcal{G}}(H) \mid H \in \mathcal{S} \}.
\label{eq:equivgateset}
\end{equation}
We define the corresponding symmetry-preserving generator set as,
\begin{equation}
    \mathcal{W}_\mathcal{G} = \{e^{-i\theta H}\mid H \in \mathcal{T}_{\mathcal{G}}(\mathcal{S})\}.
\end{equation}
Enforcing $\mathcal{G}$-symmetry typically reduces the number of independent generators (and hence trainable parameters), because twirling projects $\mathrm{span}(\mathcal{S})$ onto a subspace of generators commuting with $U_g$ for all $g \in \mathcal{G}$.
Moreover, symmetry-preserving combinations of operators often share the same trainable parameters across multiple qubits, further reducing the number of free parameters.

\subsection{Quantum Physics-informed neural network }
PINNs are a family of neural-network-based methods for solving PDEs by embedding physical constraints directly into the training objective. 
In this framework, the solution  $u(\bm{z})$  is approximated by a neural network $\hat{u}_\theta(\bm{z})$, where $\bm{z} = (\bm{x},t)$ denotes the space-time coordinates. 
A key idea is that the temporal and spatial derivatives of $\hat{u}_\theta(\bm{z})$ can be computed via automatic differentiation.
Training minimizes a loss $L(\hat{u}_{\bm{\theta}})$ that penalizes the PDE residual together with violations of boundary and initial conditions at sampled collocation points, which enables a mesh-free approximation without mesh generation.
QPINNs extend PINNs by replacing the classical neural network with a trainable quantum circuit~\cite{Siegl2025PIQC,kyriienko2021PIQC,klement2026PIQC}.

\begin{figure*}[t]
\centering
\begin{quantikz}[column sep=0.38cm]
\lstick{$q_1$} 
  & \qw
  & \gate{R_X(\theta_1)} 
  \gategroup[4,steps=8,style={dashed,rounded
    corners,fill=blue!20, inner
    xsep=2pt},background,label style={label
    position=below,anchor=north,yshift=-0.2cm}]{{\sc
    training block}}
  & \gate{R_Y(\theta_2)} 
  & \gate{R_Z(\theta_3)} 
  & \ctrl{1} 
  & \qw
  & \qw
  & \qw
  & \targ{} 
  & \gate{R_Y(x_1)} 
  \gategroup[4,steps=1,style={dashed,rounded
    corners,fill=blue!20, inner
    xsep=2pt},background,label style={label
    position=below,anchor=north,yshift=-0.2cm}]{{\sc
    encoding block}}  
  & \qw
  \slice{} 
  & \push{\,\cdots\,} 
  & \qw
  \slice{} 
  & \qw
  & \gate[4]{W_{p+1}}
  \gategroup[4,steps=1,style={dashed,rounded
    corners,fill=blue!20, inner
    xsep=2pt},background,label style={label
    position=below,anchor=north,yshift=-0.2cm}]{{\sc
    training block}}  
  & \qw
  & \meter{} \\
\lstick{$q_2$}  & \qw    
  & \gate{R_X(\theta_4)}  
  & \gate{R_Y(\theta_5)} 
  & \gate{R_Z(\theta_6)} 
  & \targ{} 
  & \qw
  & \ctrl{1} 
  & \qw
  & \qw
  & \gate{R_Y(x_2)} 
  & \qw
  & \push{\,\cdots\,} 
  &  &  &  & 
  & \meter{} \\
\wave&&&&&&&&\ctrl{1}&&\qw&\qw&\qw&\qw&\qw&\qw&\qw&\qw\\
\lstick{$q_{n}$}&\qw 
  & \gate{R_X(\theta_{3n-2})}
  & \gate{R_Y(\theta_{3n-1})}
  & \gate{R_Z(\theta_{3n})} 
  & 
  & \qw
  & \qw
  & \targ{}
  & \ctrl{-3} 
  & \gate{R_Y(t)} 
  & \qw
  & \push{\,\cdots\,} 
  & \qw
  &
  & \push{\,\cdots\,} 
  &    
  & \meter{}  
\end{quantikz}
\caption{Baseline QPINN ansatz based on the data re-uploading scheme~\cite{Siegl2025PIQC}. The circuit alternates trainable blocks and encoding blocks applied to inputs $(x_1,\ldots,x_{n-1},t)$, repeated for $p$ layers, followed by a final trainable block $W_{p+1}$ and measurement to produce the model output $u_\theta(\bm{x},t)$.}
\label{fig:conv_pqc}
\end{figure*}
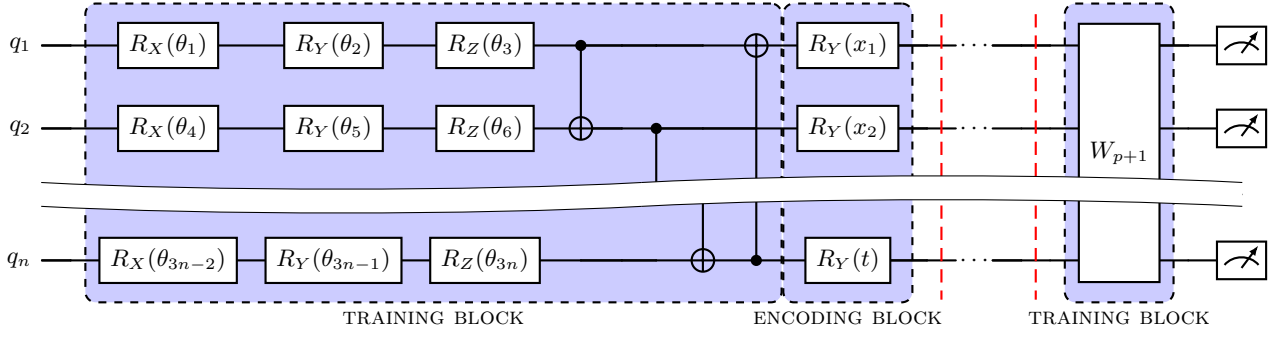

In QPINNs, the PDE solution is represented by the expectation value of a chosen observable in a parameterized quantum circuit:
\begin{equation}
    u_{\bm{\theta}}(\bm{x}, t) = \expval{O}{\psi_{\bm{\theta}}(\bm{x}, t)}.
\end{equation}
Throughout this subsection, we follow the baseline QPINN architecture in Ref.~\cite{Siegl2025PIQC}, illustrated in Fig.~\ref{fig:conv_pqc}.
We encode space-time coordinates $(\bm{x},t)$ using a data-encoding block; for example, an input scalar $s$ can be encoded as $R_y(s) = e^{-i\frac{s}{2}Y}$.
In the trainable block, single-qubit rotations generated by $\{X,Y,Z\}$ are applied independently to each qubit, followed by a ring of CNOT gates to induce entanglement, which can increase the circuit expressivity. The ansatz is composed of repeated blocks consisting of a trainable block followed by a data-encoding block, with an additional final trainable block appended at the end. We use the initial state $\ket{0}^{\otimes n}$ and choose the observable $O=\sum_{i=1}^{n} Z_i$.

Here, we consider a PDE with initial and boundary conditions:
\begin{equation}
    \begin{cases}
        &\mathcal{F}\qty[u](\bm{x},t) = 0, \ (\bm{x}, t) \in \Omega\times (0,T]\\
        &\mathcal{I}[u](\bm{x}) = 0,\ \bm{x} \in \Omega\\
        &\mathcal{B}[u](\bm{x},t) = 0,\ (\bm{x},t) \in \partial \Omega \times (0,T].
    \end{cases}
    \label{eq:PDE_setting}
\end{equation}
$\Omega$ and $\partial \Omega$ denote the spatial domain and its boundary, and $T$ denotes the final time.
$\mathcal{F}$ represents a (possibly nonlinear) differential operator that defines PDE.
$\mathcal{I}$ and $\mathcal{B}$ are the residual operators that quantify violations of the initial and boundary conditions.
In general, a PDE that is $k$th order in time requires $k$ initial conditions.

Then, the QPINN parameters $\bm{\theta}$ are trained by minimizing a physics-informed loss,
\begin{equation}
L(u_{\bm{\theta}}) = L_{\text{Res}} + \lambda_{\text{I}} L_{\text{I}} + \lambda_{\text{B}} L_{\text{B}}.
\label{QPINN_loss}
\end{equation}
Here, $L_{\text{Res}}, L_{\text{I}}$, and $L_{\text{B}}$ are the residual loss, the initial condition loss, and the boundary condition loss, respectively.
The residual loss is defined as:
\begin{equation}
L_{\text{Res}} = \frac{1}{N_{\text{Res}}} \sum_{i=1}^{N_{\text{Res}}} \left(\mathcal{F}[u_{\bm{\theta}}](\bm{x}_i, t_i) \right)^2.
\end{equation}
The residual loss $L_{\text{Res}}$ enforces the PDE structure at spatiotemporal collocation points $(\bm{x}_i, t_i) \in \Omega \times (0,T]$.
The initial condition loss $L_{\text{I}}$ is given by
\begin{equation}
L_{\text{I}} = \frac{1}{N_{\text{I}}} \sum_{i=1}^{N_{\text{I}}} (\mathcal{I}[u_{\bm{\theta}}](\bm{x}_i, 0))^2,
\end{equation}
while the boundary condition loss $L_{\text{B}}$ is expressed as
\begin{equation}
L_{\text{B}} = \frac{1}{N_{\text{B}}} \sum_{i=1}^{N_{\text{B}}} (\mathcal{B}[u_{\bm{\theta}}](\bm{x}_i, t_i))^2.
\end{equation}
Here, $N_\mathrm{Res}, N_{\mathrm{I}}$ and $N_{\mathrm{B}}$ denote the number of collocation points used for the residual loss, the initial condition loss, and the boundary condition loss.
Minimizing $L(u_{\bm{\theta}})$ enforces the PDE together with the initial and boundary conditions.
Thus, $u_{\bm{\theta}}$ provides a good approximation to the PDE solution under the specified initial and boundary conditions.

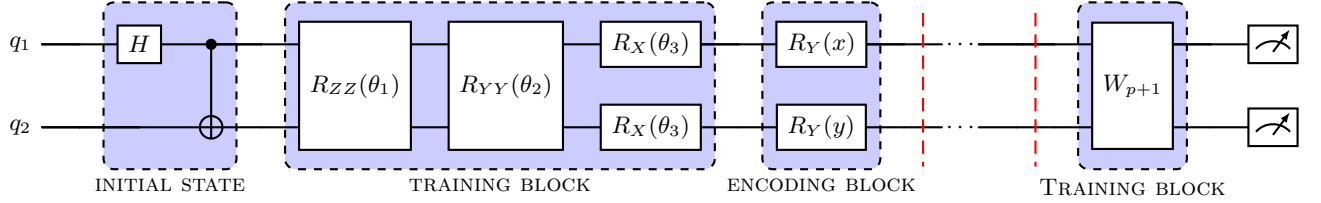
\begin{figure*}[t]
\centering
\begin{quantikz}
\lstick{$q_1$} & \qw  & \gate{H} \gategroup[2,steps=2,style={dashed,rounded
corners,fill=blue!20, inner
xsep=2pt},background,label style={label
position=below,anchor=north,yshift=-0.2cm}]{{\sc
initial state}}& \ctrl{1} 
  & \qw
  & \gate[2]{R_{ZZ}(\theta_1)} 
  \gategroup[2,steps=3,style={dashed,rounded
    corners,fill=blue!20, inner
    xsep=2pt},background,label style={label
    position=below,anchor=north,yshift=-0.2cm}]{{\sc
    training block}}
  & \gate[2]{R_{YY}(\theta_2)} 
  & \gate{R_X(\theta_3)} 
  & \qw
  & \gate{R_Y(x)} 
  \gategroup[2,steps=1,style={dashed,rounded
    corners,fill=blue!20, inner
    xsep=2pt},background,label style={label
    position=below,anchor=north,yshift=-0.2cm}]{{\sc
    encoding block}}  
  & \qw
  \slice{} 
  & \push{\,\cdots\,} 
  & \qw
  \slice{} 
  & \qw
  & \gate[2]{W_{p+1}}
  \gategroup[2,steps=1,style={dashed,rounded
    corners,fill=blue!20, inner
    xsep=2pt},background,label style={label
    position=below,anchor=north,yshift=-0.2cm}]{{\sc
    Training block}}  
  & \qw
  & \meter{} \\
\lstick{$q_2$} & \qw & \qw     & \targ{} 
  & & &
  & \gate{R_X(\theta_3)} &
  & \gate{R_Y(y)} &
  & \push{\,\cdots\,} 
  &  &  &  & 
  & \meter{}
\end{quantikz}
\caption{$K_4$-GQPINN ansatz enforcing invariance under coordinate exchange $(x,y)\mapsto(y,x)$ and simultaneous parity inversion $(x,y)\mapsto(-x,-y)$. The trainable block is constructed to be $K_4$-equivariant using the symmetry-preserving generator set $\mathcal{T}_{K_4}$. The coordinates $(x,y)$ are encoded onto the two qubits via angle encoding using $R_Y$ rotations.}
\label{fig:sym_circuit}
\end{figure*}

\section{Theoretical Framework for Symmetry-Embedded Circuit Design}
\label{sec:group_rep}

\begin{figure*}[t]
\centering
\begin{quantikz}
\lstick{$q_1$} & \qw  & \gate{H} \gategroup[2,steps=2,style={dashed,rounded
corners,fill=blue!20, inner
xsep=2pt},background,label style={label
position=below,anchor=north,yshift=-0.2cm}]{{\sc
initial state}}  & \ctrl{1} 
  & \qw
  & \gate[2]{R_{ZZ}(\theta_1)} 
  \gategroup[2,steps=3,style={dashed,rounded
    corners,fill=blue!20, inner
    xsep=2pt},background,label style={label
    position=below,anchor=north,yshift=-0.2cm}]{{\sc
    training block}}
  & \gate[2]{R_{XX+YY}(\theta_2)} 
  & \gate{R_Z(\theta_3)} 
  & \qw
  & \gate{\tilde{U}(x,y)}  
  \gategroup[2,steps=1,style={dashed,rounded
    corners,fill=blue!20, inner
    xsep=2pt},background,label style={label
    position=below,anchor=north,yshift=-0.2cm}]{{\sc
    encoding block}}  
  & \qw
  \slice{} 
  & \push{\,\cdots\,} 
  & \qw
  \slice{} 
  & \qw
  & \gate[2]{W_{p+1}}
  \gategroup[2,steps=1,style={dashed,rounded
    corners,fill=blue!20, inner
    xsep=2pt},background,label style={label
    position=below,anchor=north,yshift=-0.2cm}]{{\sc
    training block}}  
  & \qw
  & \meter{} \\
\lstick{$q_2$} & \qw & \gate{X}   & \targ{} 
  & & &
  & \gate{R_Z(\theta_4)} &
  & \gate{\tilde{U}(x,y)} &
  & \push{\,\cdots\,} 
  &  &  &  & 
  & \meter{}
\end{quantikz}
\caption{$SO(2)$-GQPINN ansatz enforcing invariance under rotations in the $(x,y)$ plane. The overall circuit structure follows Fig.~\ref{fig:sym_circuit}, with the trainable block constructed from the symmetry-preserving generator set $\mathcal{T}_{SO(2)}$. The coordinates $(x,y)$ are encoded onto the qubits using the Bloch-sphere encoding unitary $\tilde{U}(x,y)$ in Eq.~\eqref{eq:bloch_encode}. }
\label{fig:cont_sym_circuit}
\end{figure*}
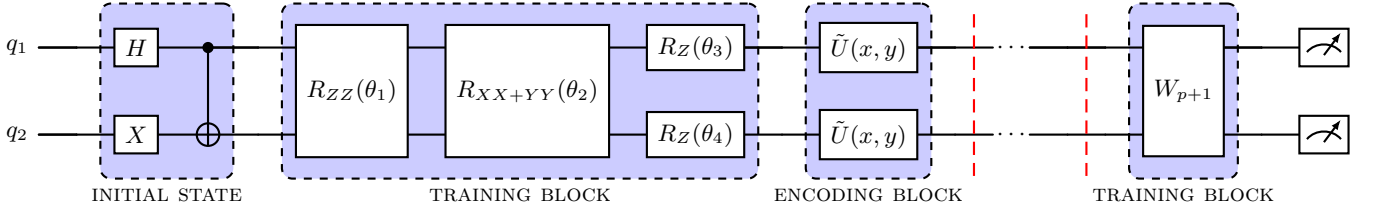

\subsection{Symmetry in QPINN}

As shown in Eq.~\eqref{eq:PDE_setting}, a PDE solution is determined not only by the governing equation but also by the prescribed initial and boundary conditions.  
Even when a transformation preserves the governing PDE, it may either leave the prescribed initial and boundary conditions unchanged or transform them into different ones.  
Let $\mathcal{G}$ be a group of transformations that preserve the governing PDE, acting on the space-time coordinates $\bm{z}=(\bm{x},t)$ through transformations $V_g$, $g\in\mathcal{G}$.
We also allow the group to act on the output space of the solution through a representation $K_g$.  
The induced action on functions is defined by
\begin{equation}
    (T_g u)(\bm{z}) := K_g u(V_g^{-1}[\bm{z}]).
    \label{eq:induced_action}
\end{equation}
We restrict attention to transformations that preserve the spatial domain $\Omega$, its boundary $\partial\Omega$, and the initial-time slice $t=0$.

Let $d$ denote the prescribed initial and boundary conditions, and let
\begin{equation}
    \mathcal{S}_d = \{u:\mathcal{F}[u]=0,\ \mathcal{I}_d[u]=0,\ \mathcal{B}_d[u]=0\}
\end{equation}
be the corresponding solution set. 
A transformation is a symmetry of the fixed initial-boundary value problem with data $d$ if
\begin{equation}
    T_g \mathcal{S}_d = \mathcal{S}_d.
    \label{eq:full_problem_symmetry}
\end{equation}
Here, $T_g\mathcal{S}_d = \{T_g u:u\in\mathcal{S}_d\}$.
In this case, the transformed solution satisfies the same governing PDE together with the same initial and boundary conditions.  

By contrast, a transformation may preserve the governing PDE while changing
the prescribed initial or boundary conditions.  
In that case,
\begin{equation}
    T_g \mathcal{S}_d = \mathcal{S}_{g\cdot d},
    \label{eq:family_equivariance}
\end{equation}
where $g\cdot d$ denotes the transformed initial and boundary conditions, including both the coordinate action $V_g$ and the output-space action $K_g$.
Such transformations relate different problem instances rather than leaving a single fixed initial-boundary value problem invariant.  

This distinction is essential for PINNs and QPINNs. 
In their standard formulation, the network is trained with fixed initial and boundary conditions to approximate a single solution $u_d$.  
Consequently, as a symmetry constraint on the learned solution, a standard single-instance PINN/QPINN can directly impose only transformations satisfying Eq.~\eqref{eq:full_problem_symmetry}.
In the following, we therefore restrict our discussion to transformations that leave the full fixed problem invariant and, for simplicity, continue to denote the corresponding subgroup by $\mathcal{G}$.

If the fixed initial-boundary value problem has a unique solution $u_d$, then Eq.~\eqref{eq:full_problem_symmetry} implies
\begin{equation}
    T_g u_d = u_d .
    \label{eq:unique_solution_symmetry}
\end{equation}
Using Eq.~\eqref{eq:induced_action}, the corresponding constraint on a PINN/QPINN approximation $u_\theta$ can be written as
\begin{equation}
    u_\theta(V_g[\bm{z}]) = K_g u_\theta(\bm{z}).
    \label{eq:solution_equivariance}
\end{equation}
Here, the equivariance in Eq.~\eqref{eq:solution_equivariance} is a solution-level equivariance induced by the fixed-problem symmetry in Eq.~\eqref{eq:full_problem_symmetry}, and should be distinguished from the problem-family equivariance in Eq.~\eqref{eq:family_equivariance}.
This is precisely the \(\mathcal{G}\)-equivariance condition in
Eq.~\eqref{eq:equivariance_exp}.
Thus, to impose this symmetry, it suffices to design the QPINN ansatz so that its output satisfies the corresponding equivariance.
When $K_g=\mathrm{Id}$ for all $g\in\mathcal{G}$, Eq.~\eqref{eq:solution_equivariance} reduces to a $\mathcal{G}$-invariant model.

In this work, we focus on spatial symmetries and consider transformations that leave time unchanged,
\begin{equation}
    V_g[(\bm{x},t)] = (\tilde{V}_g[\bm{x}],t).
\end{equation}
We then design the QPINN ansatz so that its output satisfies
\begin{equation}
    u_\theta(\tilde{V}_g[\bm{x}],t) = K_g u_\theta(\bm{x},t),
    \ \forall \bm{x}\in\Omega,\ t\in(0,T],\ g\in\mathcal{G}.
    \label{eq:gqpinn_equivariance}
\end{equation}
In this study, we focus on scalar-valued output represented by a single observable $O$.
Hence, $K_g$ is understood as scalar multiplication given by a one-dimensional representation, and Eq.~\eqref{eq:gqpinn_equivariance} becomes
\begin{equation}
\begin{split}
    \expval{O}{\psi_{\bm{\theta}}(\tilde{V}_g[\bm{x}],t)}
    &=
    K_g \expval{O}{\psi_{\bm{\theta}}(\bm{x},t)}
    \\
    &\forall \bm{x}\in\Omega,\ t\in(0,T],\ g\in\mathcal{G}.
\end{split}
    \label{eq:equi_space_eq}
\end{equation}
We refer to such a QPINN ansatz as a \emph{geometric QPINN (GQPINN) ansatz}.

Following the GQML framework adopted here, we restrict attention to transformations associated with finite groups and compact Lie groups. 
While every finite group is compact in the discrete topology, a continuous symmetry group need not be compact. Accordingly, our GQPINN framework can accommodate both finite discrete groups and compact Lie groups. In this work, however, we focus on solution invariances induced by the actions of the Klein four group $K_4$ and compact rotation group $SO(2)$, as representative examples of a finite discrete group and a compact Lie group, respectively.
We discuss these invariant models in the following subsections.
We also discuss an equivariant model in Sec.~\ref{app:Z_2_case}.

\subsection{Representation of Klein Four-Group $K_4$}
\label{sec:discrete}
In this subsection, we derive a $K_4$-GQPINN ansatz for solutions that are invariant under Klein's four-group.
These symmetries arise naturally for the two-dimensional Poisson problem with homogeneous Dirichlet boundary conditions whenever the domain and forcing term are invariant under the same transformations.
Suppose the solution $ u(x, y)$ is invariant under coordinate exchange $(x, y) \to (y, x)$ and parity inversion $(x, y) \to (-x, -y)$. 
Consequently, it satisfies the symmetry relations
\begin{equation}
u(x, y) = u(y, x), \quad u(x, y) = u(-x, -y).
\end{equation}
Accordingly, we take the symmetry group to be $K_4=\{e,s,p,sp\}$, which is isomorphic to $\mathbb{Z}_2 \times \mathbb{Z}_2$.
The representation of each element can be written as
\begin{equation}
\begin{aligned}
&V_e(x,y) = (x,y), \quad V_p(x,y) = (-x,-y), \\
&V_s(x,y) = (y,x), \quad V_{sp}(x,y) = (-y,-x).
\end{aligned}
\label{eq:group_k4}
\end{equation}

For simplicity, we consider a two-qubit circuit. 
Each component of the input $(x, y) \in \mathbb{R}^2$ is individually encoded onto a qubit using a parameterized rotation $R_Y(\theta) = e^{-i \theta Y /2}$.
The data-encoding block is given by
\begin{equation}
U(x, y) = R_Y(x) \otimes R_Y(y).
\label{eq:Ry_encoding}
\end{equation}
Based on this encoding block, the corresponding induced representations are given by
\begin{equation}
\begin{aligned}
U_{e} &= I \otimes I, \ &
U_{p} &= X \otimes X, \\
U_{s} &= \text{SWAP}, \quad &
U_{sp} &= \text{SWAP} \cdot (X \otimes X).
\end{aligned}
\label{eq:induced_rep}
\end{equation}

In order to construct a $K_4$-invariant re-uploading model, we choose the trainable block from the commuting generator set $\mathcal{T}_{K_4}(\mathcal{S})$, whose elements commute with $U_g$ for all $g \in K_4$.
To begin with, we consider a generator set, consisting of single-qubit Pauli terms $\{X,Y,Z\}$, and two-qubit interaction terms $\{XX, YY,ZZ\}$. 
The resulting generator set can be written as
\begin{equation}
    \mathcal{S} = \{ X_1, Y_1, Z_1, X_2, Y_2, Z_2, X_1 X_2, Y_1 Y_2, Z_1 Z_2 \}.
\label{eq:gateset}
\end{equation}
The commuting generator set for $K_4$ group can be computed using twirling formula Eq.~\eqref{eq:twirling}, incorporating unitary representation in Eq.~\eqref{eq:induced_rep}.
Then, we obtain the commuting generator set for $K_4$ group as
\begin{equation}
    \mathcal{T}_{K_4}(\mathcal{S}) = \bigg\{ \frac{X_1 + X_2}{2},\ X_1 X_2,\ Y_1 Y_2,\ Z_1 Z_2 \bigg\}.
    \label{eq:klein_gates}
\end{equation}
We observe that $X_1$ and $X_2$ are symmetrized to the same averaged generator $(X_1 + X_2)/2$, while $\{Y_1,Y_2, Z_1, Z_2\}$ are eliminated by the twirling procedure.
As a result, symmetrization reduces the number of independent generators (and hence trainable parameters), which in turn simplifies the circuit.

We also need an invariant initial state and an invariant observable to construct this model.
To satisfy Eq.~\eqref{eq:state-invariance}, we prepare the Bell state $\ket{\Phi^+} = \frac{1}{\sqrt{2}} (\ket{00} + \ket{11})$ as the initial state and choose the observable as $O = X_1 + X_2$. 
Both the state and the observable remain invariant under all unitary representations $U_{g}\ (g \in K_4)$ defined in Eq.~\eqref{eq:induced_rep}.
Figure~\ref{fig:sym_circuit} illustrates the resulting $K_4$- GQPINN ansatz.

\subsection{Representation of Rotational Symmetry $SO(2)$}
We consider rotational symmetry in the $(x,y)$-plane and construct an $SO(2)$-GQPINN ansatz. Since $SO(2)$ is an infinite, compact, one-parameter Lie group, it admits a normalized Haar measure and unitary representations on an appropriate Hilbert space. In the following, we use such a unitary representation to embed rotational invariance into the QPINN circuit design, ensuring that the model output is invariant under planar rotations.

Leaving $V_\varphi$ denote a representation of rotation around the $z$-axis, we impose
\begin{equation}
  u(x, y) = u\qty(V_\varphi  (x,  y)),\qquad \varphi \in [0, 2\pi).
\end{equation}
where the rotation operator $V_\varphi$ is defined by
\begin{equation}
V_\varphi  = \begin{bmatrix}\cos\varphi & -\sin\varphi \\\sin\varphi & \cos\varphi \end{bmatrix}.
\label{eq:rotation_op}
\end{equation}

Now, we consider a data-encoding block that maps classical data points $(x, y)$ onto the Bloch sphere,
\begin{equation}
    \tilde{U}(x,y) = e^{-\frac{i}{2}(x X + y Y)}.
\label{eq:bloch_encode}
\end{equation}
We can derive the induced representation of $SO(2)$ as
\begin{equation}
    \begin{aligned}
        \tilde{U}(V_\varphi (x, y))
        &=R_Z(\varphi)e^{-\frac{i}{2}(xX+yY) }R_Z(\varphi)^{\dagger}\\
        &=R_Z(\varphi) \tilde{U}(x,y) R_Z(\varphi)^{\dagger},
    \end{aligned}
\end{equation}
where $R_Z(\varphi) = e^{-i\varphi Z/2}$.
Thus, the induced unitary representation corresponds to a Pauli $Z$-rotation, and we denote the set of induced representation as
\begin{equation}
    U_{SO(2)}  = \left\{ R_Z(\varphi) \;\middle|\; \varphi \in [0, 2\pi) \right\}.
\end{equation}

Since $SO(2)$ is a continuous group, the representation $U_{SO(2)}$ is indexed by infinitely many group elements.
Consequently, the twirling operation in Eq.~\eqref{eq:twirling_cont} is expressed as a Haar integral rather than a finite sum and can be written as
\begin{equation}
    T_{SO(2)}(H)=\frac{1}{2\pi}\int_0^{2\pi} R_Z(\varphi) H R_Z^\dagger(\varphi) \, d\varphi.
    \label{eq: twirling_SO2}
\end{equation}

We chose the two-qubit embedding to maintain a fair comparison with the $K_4$-GQPINN ansatz, which also operates on a two-qubit Hilbert space. For the $SO(2)$-GQPINN ansatz, this choice additionally places the model in a quantum-enhanced setting~\cite{ragone2022GQML, huang2022quantumadvantage, larocca2023groupequivariant, Nguyen2024equivariantQC}, since the same input is available in two copies and can be processed jointly.
We can obtain the equivariant generator set  $\mathcal{T}_{SO(2)}[\mathcal{S}]$ using Eq.~\eqref{eq: twirling_SO2}, and the resulting generator set  is
\begin{equation}
    \mathcal{T}_{SO(2)}[\mathcal{S}] = \bigg\{ Z_1, Z_2,\ Z_1 Z_2,\ \frac{1}{2}(X_1X_2+Y_1Y_2) \bigg\}.
\label{eq:cont_sym_gateset}
\end{equation}
Both $X_1X_2$ and $Y_1Y_2$ are mapped to $\frac{1}{2}(X_1X_2+Y_1Y_2)$, and $X_i$ and $Y_i$ do not contribute to the set of equivariant gates.

\begin{figure*}[t]
\centering
\begin{quantikz}[column sep=0.21cm]
\lstick{$q_1$} & \qw  & \gate{H} \gategroup[3,steps=2,style={dashed,rounded
corners,fill=blue!20, inner
xsep=1pt},background,label style={label
position=below,anchor=north,yshift=-0.2cm}]{{\sc
initial state}}  & \ctrl{1} 
  & \qw
  & \gate{R_Z(\theta_1)}  
  \gategroup[3,steps=8,style={dashed,rounded
    corners,fill=blue!20, inner
    xsep=2pt},background,label style={label
    position=below,anchor=north,yshift=-0.2cm}]{{\sc
    training block}}
  & \gate[2]{R_{XX+YY}(\theta_3)} 
  & \gate[2]{R_{ZZ}(\theta_4)} 
  & 
  &  \gate[3]{R_{ZZ}(\theta_8)} 
  & \qw
  & 
  & 
  & \qw
  & \gate{\tilde{U}(x,y)}  
  \gategroup[3,steps=1,style={dashed,rounded
    corners,fill=blue!20, inner
    xsep=0.5pt},background,label style={label
    position=below,anchor=north,yshift=-0.2cm}]{{\sc
    encoding block}}  
  & \qw
  \slice{} 
  & \push{\,\cdots\,} 
  \slice{} 
  & \qw
  & \qw
  & \gate[3]{W_{p+1}}
  \gategroup[3,steps=1,style={dashed,rounded
    corners,fill=blue!20, inner
    xsep=1pt},background,label style={label
    position=below,anchor=north,yshift=-0.2cm}]{{\sc}}  
  & \qw
  & \meter{} \\
\lstick{$q_2$} & \qw & \gate{X}   & \targ{} 
  & & \gate{R_Z(\theta_2)}  & &
  &  
  &
  & 
  & \qw
  &  \gate[2]{R_{ZZ}(\theta_9)} 
  & 
  & \gate{\tilde{U}(x,y)} 
  & 
  & \push{\,\cdots\,} 
  &  &  &  & 
  & \meter{}\\
\lstick{$q_3$} & \qw & \qw  & \qw
  & & \gate{R_Z(\theta_5)}& \gate{R_Y(\theta_6)}
  & \gate{R_X(\theta_7)} 
  & 
  &
  & \qw
  & \qw
  & \qw
  & \qw
  & \gate{R_Z(t)} &
  & \push{\,\cdots\,} 
  &  &  &  & 
  & \meter{}
\end{quantikz}
\caption{The extension of $SO(2)$-GQPINN ansatz for time-dependent PDEs. The spatial qubits follows the $SO(2)$-GQPINN structure in Fig.~\ref{fig:cont_sym_circuit}, while the time qubit follows the baseline QPINN structure in Fig.~\ref{fig:conv_pqc} with time encoded via $R_Z(t)$. The only additional components are two $R_{ZZ}$ couplings, parameterized by $\theta_8$ and $\theta_9$,that entangle the spatial and temporal registers to enable spatiotemporal interactions.}
\label{fig:diffusion_circuit}
\end{figure*}
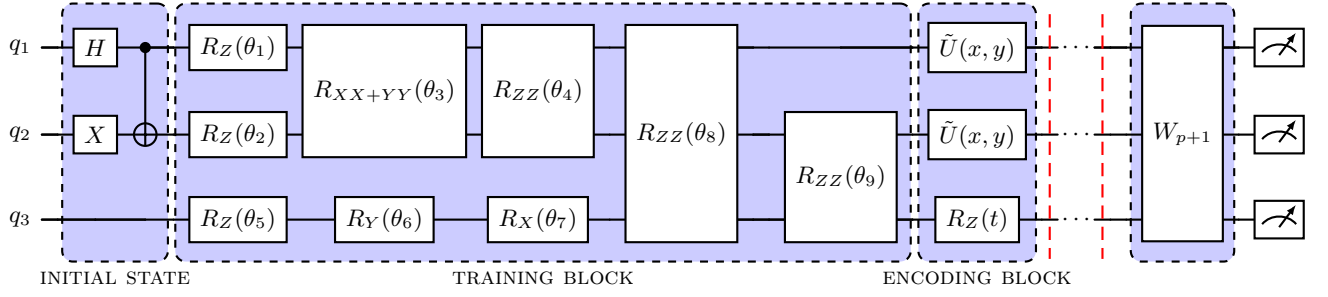

For the invariant initial state and the invariant observable, we prepare the Bell state $\ket{\Psi^+} = \frac{1}{\sqrt{2}} (\ket{01} + \ket{10})$ as the initial state and choose the observable to be $O=\frac{1}{2}(X_1X_2+Y_1Y_2)$. 
We illustrated the complete $SO(2)$-GQPINN ansatz in Fig.~\ref{fig:cont_sym_circuit}. 

\subsection{Extension to Time-Dependent Dynamics}
When spatial symmetry exists, our framework naturally extends to time-dependent PDEs by retaining the symmetry-enforcing spatial ansatz and augmenting it with a temporal encoding. 
Spatial coordinates $(x,y)$ are encoded using the scheme in Eq.~\eqref{eq:bloch_encode}, while the temporal variable $t$ is incorporated via an angle-encoding rotation gate $R_Z(t)$ acting on an additional qubit. 
Consistently, the equivariant generator set  $\mathcal{T}_{SO(2)}[\mathcal{S}]$ acts only on the spatial qubits, whereas the temporal qubit is governed by the unrestricted generator set  $\mathcal{S}$ Eq.~\eqref{eq:gateset}. Entanglement between the spatial and temporal qubits is essential, since otherwise the temporal qubit would remain decoupled from the symmetry-adapted features encoded in the spatial subsystem. Within the equivariant generator set  $\mathcal{T}_{SO(2)}[\mathcal{S}]$, only the spatial $Z$ component remains invariant, whereas the spatial $X$ and $Y$ components average to zero under the group action. Consequently, symmetry-compatible entanglement between the spatial and temporal qubits must be mediated by couplings involving Pauli $Z$ on the spatial qubits, which motivates the use of $R_{ZZ}$.
The initial state measurement observable are set to be $\frac{1}{\sqrt{2}} (\ket{01} + \ket{10}) \otimes \ket{0}$ and $\frac{1}{2}(X_1X_2+Y_1Y_2)+ Z_3$ respectively. 
The complete extension of $SO(2)$-GQPINN ansatz for time-dependent PDEs is illustrated in Fig.~\ref{fig:diffusion_circuit}. 

\section{Experimental Results on Poisson and Diffusion Equations}
\begin{table}
\caption{Hyperparameters setting of the L-BFGS optimizer}
\label{tab:hyperparams}
\centering
\renewcommand{\arraystretch}{1.2}
\begin{tabular}{cc}
\hline\hline
 & \textbf{2D Poisson Eq.}  \\
\hline
\texttt{lr}                     & 0.7     \\
\texttt{max\_iter}             & 20   \\
\texttt{max\_eval}             & 25   \\
\texttt{tolerance\_grad}       & $10^{-7}$ \\
\texttt{tolerance\_change}     & $10^{-9}$ \\
\texttt{history\_size}         & 100  \\
\texttt{line\_search\_fn}      & \texttt{strong\_wolfe} \\
\hline\hline
\end{tabular}
\end{table}

\label{sec:experimental_result}
In this section, we examine the performance of the GQPINN ansatzes developed in the previous section and compared with the QPINN ansatz.
First, in Section ~\ref{sec:poisson}, we compare the performance of QPINN, $K_4$-GQPINN ,and $SO(2)$-GQPINN for the Poisson equation. 
In Section ~\ref{sec:diffusion}, we evaluate the performance of the extension $SO(2)$-GQPINN for the diffusion equation.

We employ the same optimizer used in the previous study~\cite{Siegl2025PIQC}, the Limited-memory Broyden-Fletcher-Goldfarb-Shanno quasi-Newton algorithm (L-BFGS)~\cite{Liu1989LBFGS}. 
The optimizer’s parameter settings are provided in Table~\ref{tab:hyperparams}. All training runs are performed for 50 epochs, repeated over 10 independent samples. To evaluate the quality of the training, we use mean absolute error (MAE), defined as 
\begin{equation}
\text{MAE} = \frac{1}{N_\mathrm{val}} \sum_{i=1}^{N_{val}} \abs{ \hat{u}_\theta - u_{\text{exact}} },
\label{eq:mae}
\end{equation}
where $N_\mathrm{val}$ is the number of validation points. 
All quantum algorithms are implemented using Pennylane~\cite{Bergholm2018pennylane}, and all quantum circuit simulations were performed by using statevector method on an Apple M2 chip.

\subsection{2D Poisson Equation}
\label{sec:poisson}

\begin{figure}[t]
\centering
\includegraphics[width=\linewidth]{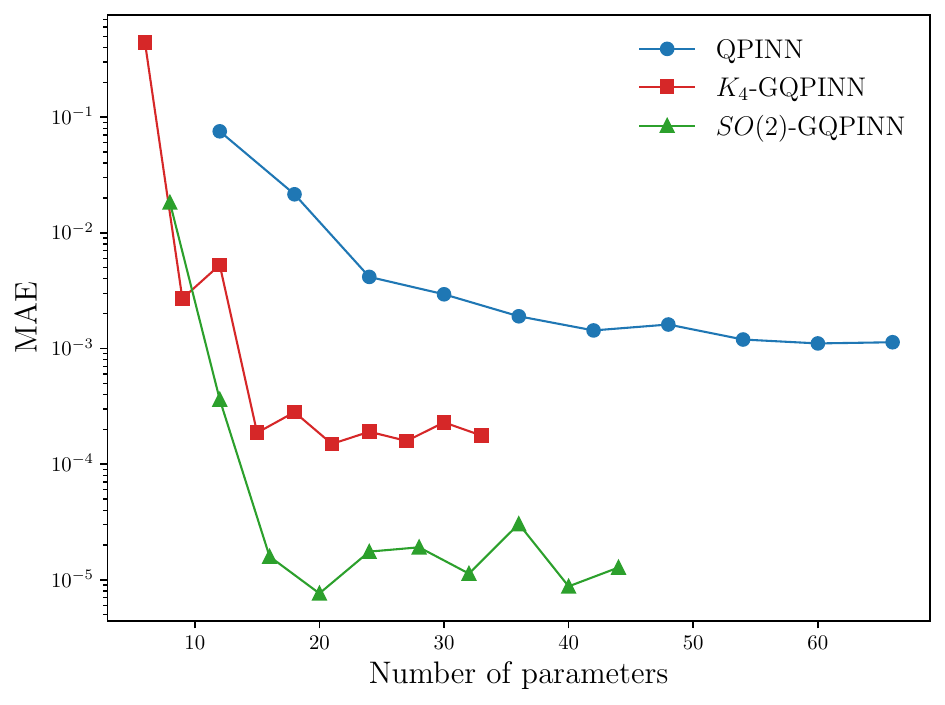} 
\caption{MAE of the QPINN, $K_4$-GQPINN, and $SO(2)$-GQPINN ansatzes on the 2D Poisson equation across different numbers of trainable parameters.}
\label{fig:benchmark_static_param}
\end{figure}

Let us consider the two-dimensional Poisson equation with a constant source term:
\begin{equation}
\frac{\partial^2 u}{\partial x^2} + \frac{\partial^2 u}{\partial y^2} = \frac{1}{D},
\label{2dpossion}
\end{equation}
posed on a bounded domain $\Omega \subset \mathbb{R}^2$, together with homogeneous Dirichlet boundary conditions,
\begin{equation}
u(x,y) = 0, \qquad (x,y)\in \partial\Omega.
\label{eq:bc_dirichlet}
\end{equation}
Here, we used the circular domain to enforce the boundary condition satisfies the symmetry.
The diffusion coefficient is set to $D = 1$ throughout the experiments. 
In this setting, the analytical solution is $u_{\text{exact}}(x, y)=\frac{1}{4}(x^2+y^2-1)$.We sampled $N_{\text{Res}}=276$ data points from the domain 
$\{(x,y)\,|\, x^2+y^2 \leq 1\}$, and $N_{\text{B}}=100$ the boundary points from 
$\{(x,y)\,|\, x^2+y^2 = 1\}$. 
The number of trainable parameters per trainable block is 3, 4, and 6 for the $K_4$-GQPINN, $SO(2)$-GQPINN, and conventional QPINN ansatzes, respectively.

First, we compare the MAE across the three ansatz constructions while fixing the number of qubits to 2 and increasing the number of repeated layers from 1 to 10. To ensure a fair comparison despite the different parameter counts per trainable block, we report MAE as a function of the total number of trainable parameters in Fig.~\ref{fig:benchmark_static_param}.
This allows us to assess whether the GQPINN ansatzes achieves improved accuracy even with fewer parameters. 
Figure ~\ref{fig:benchmark_static_param} shows that both GQPINN ansatzes achieve substantially smaller MAE while using fewer trainable parameters. 
In our experiments, $SO(2)$-GQPINN ansatz achieved up to two orders of magnitude lower in MAE compared with the QPINN ansatz, even though the QPINN ansatz uses roughly three times as many trainable parameters.

\begin{figure}[t]
\centering
\includegraphics[width=\linewidth]{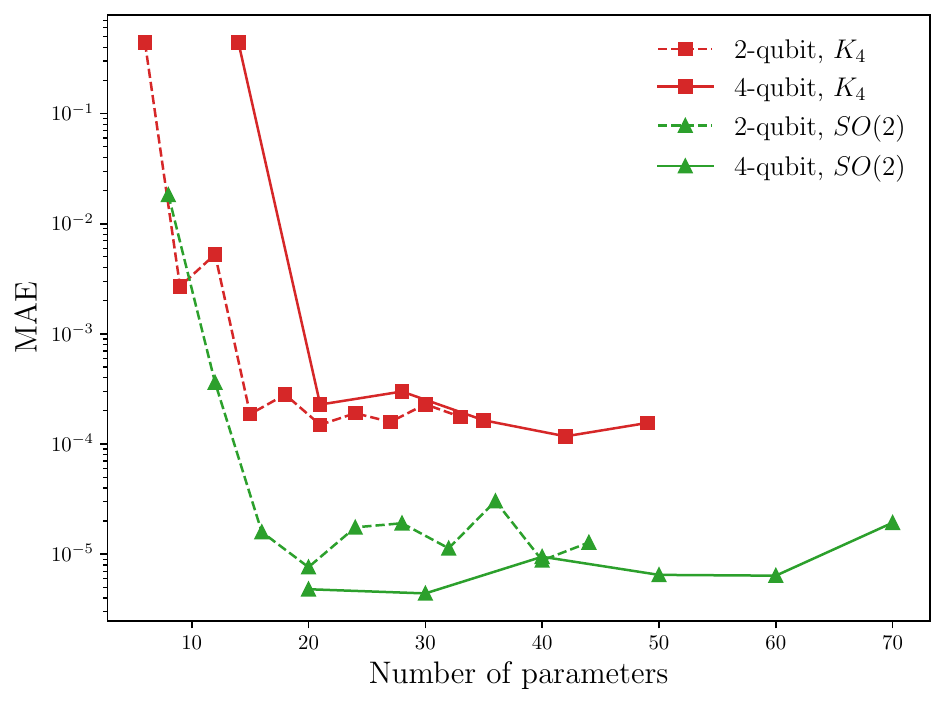} 
\caption{MAE of the $K_4$- and $SO(2)$-GQPINN models (2- and 4-qubit settings) on the 2D Poisson equation across different numbers of trainable parameters.}
\label{fig:benchmark_4qubits}
\end{figure}

Next, we evaluate the dependence on the number of qubits by increasing the qubit count to 4.
We used as generator set  as
\begin{equation}
\begin{aligned}
    \mathcal{S}^{(4)} &= \{X_1, X_2, X_3, X_4, Z_1, Z_2, Z_3, Z_4,X_1 X_2,\\
    & X_2 X_3, X_3 X_4, Y_1 Y_2, Y_3 Y_4, Z_1 Z_2, Z_2 Z_3, Z_3 Z_4 \}.
\end{aligned}
\end{equation}
and then we obtain the following equivariant generator set using twirling formula:
\begin{equation}
\begin{aligned}
\mathcal{T}_{K_4}[\mathcal{S}^{(4)}]=
\bigg\{&
\tfrac{X_1 + X_2}{2},\ \tfrac{X_3 + X_4}{2},\ 
Y_1 Y_2,\ Y_3 Y_4,\ 
Z_1 Z_2,\\
&Z_3 Z_4, \tfrac{X_2 X_3+ X_1 X_3 + X_2 X_4+ X_1 X_4}{4}
\bigg\}.\\
\mathcal{T}_{SO(2)}[\mathcal{S}^{(4)}] =
\bigg\{&
Z_1,\ Z_2,\ Z_3,\ Z_4,\ 
Z_1 Z_2,\ Z_2 Z_3,\ Z_3 Z_4,\\
&\tfrac{X_1X_2+Y_1Y_2}{2},\ 
\tfrac{X_2X_3+Y_2Y_3}{2},\
\tfrac{X_3X_4+Y_3Y_4}{2}
\bigg\}.
\end{aligned}
\end{equation}
We fix the number of layers to 6 for the 4-qubit setting and compare the MAE across the $K_4$- and $SO(2)$-GQPINN ansatzes.
We also compare the 2- and 4-qubit settings for both symmetry constructions in Fig.~\ref{fig:benchmark_4qubits}.

Increasing the number of qubits enlarges the Hilbert space and increases the number of trainable parameters per layer, so we report results as MAE versus the total parameter count. As shown in Fig.~\ref{fig:benchmark_4qubits}, for both $K_4$- and $SO(2)$-GQPINN, the 2- and 4-qubit curves approach similar MAE values once the number of parameters are comparable, indicating that the total number of trainable degrees of freedom largely determines performance in this regime.
At small parameter counts, the $K_4$- and $SO(2)$-GQPINN models can behave differently. 
For $K_4$-GQPINN, the 4-qubit model performs worse at the very beginning despite having a similar parameter count to the 2-qubit model. 
For $SO(2)$-GQPINN, the 4-qubit model appears slightly less variable at moderate parameter counts, although both settings converge to similar MAE values once the total number of trainable parameters is comparable. Overall, the plot suggests that increasing the number of qubits does not provide a systematic advantage at matched parameter counts.

\begin{figure}[]
\centering
\includegraphics[width=\linewidth]{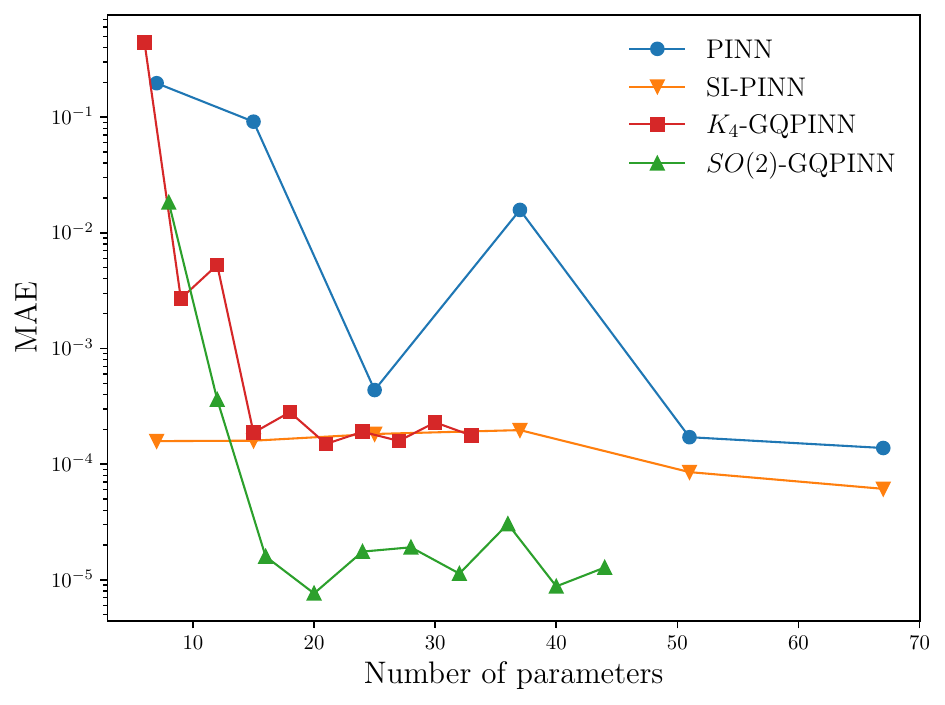} 
\caption{MAE of classical (PINN, SI-PINN) and quantum ($K_4$-GQPINN, $SO(2)$-GQPINN) models on the 2D Poisson equation across different numbers of trainable parameters.}
\label{fig:benchmark_static_pinn}
\end{figure}

Finally, we also compare the two GQPINN ansatzes against classical physics-informed neural network (PINN) baselines. Specifically, we consider two PINN models. The first is a standard PINN, serving as the classical baseline corresponding to the conventional QPINN ansatz. We additionally consider a symmetry-invariant PINN (SI-PINN) aligned with our symmetry-embedded constructions based on Klein four-group $K_4$. 
Since the classical SI-PINN construction relies on finite group averaging, it does not directly yield a finite-sum construction for exact invariance under the full continuous group $SO(2)$.  
We therefore do not consider an SI-PINN construction for the $SO(2)$-invariant case in this work.
Details of the SI-PINN constructions for both groups are deferred to Appendix~\ref{app:sinn}.
All PINN models use a single hidden layer with $\tanh$ activation functions, and the number of hidden nodes is varied from $1$ to $6$.  For a fair comparison on the 2D Poisson equation, we match the number of trainable parameters, the number of data points, and the optimizer settings across methods as closely as possible.

Figure~\ref{fig:benchmark_static_pinn} reports the MAE as a function of the number of trainable parameters. 
SI-PINN outperforms the standard PINN, indicating that incorporating symmetry is beneficial even in the classical model.
SI-PINN achieves low MAE in the small-parameter counts, although its performance does not improve further as the parameter count increases.
Comparing SI-PINN and GQPINN models, the performance gap depends on the symmetry construction. $K_4$-GQPINN is competitive with the SI-PINN but does not provide a consistent improvement at comparable parameter counts.  
Taken together, the $SO(2)$-GQPINN attains the lowest MAE overall. 
Overall, these results suggest that symmetry constraints improve parameter efficiency in both classical and quantum models, and that the $SO(2)$-GQPINN ansatz achieves the best performance in this benchmark. 
In $SO(2)$-GQPINN, the chosen data encoding maps spatial inputs to Bloch-sphere rotations, for which $SO(2)$ acts naturally as a continuous rotation symmetry. This alignment between the problem symmetry and the geometry of single-qubit rotations can make rotational invariance easier to represent by construction, which may further contribute to the strong performance of the $SO(2)$-GQPINN.
In addition, the $SO(2)$-GQPINN uses a quantum-enhanced setting, which may provide additional representational flexibility and help explain the performance gap relative to the classical SI-PINN baselines.

\subsection{2D Diffusion Equation}
\label{sec:diffusion}
\begin{figure}[]
\centering
\includegraphics[width=\linewidth]{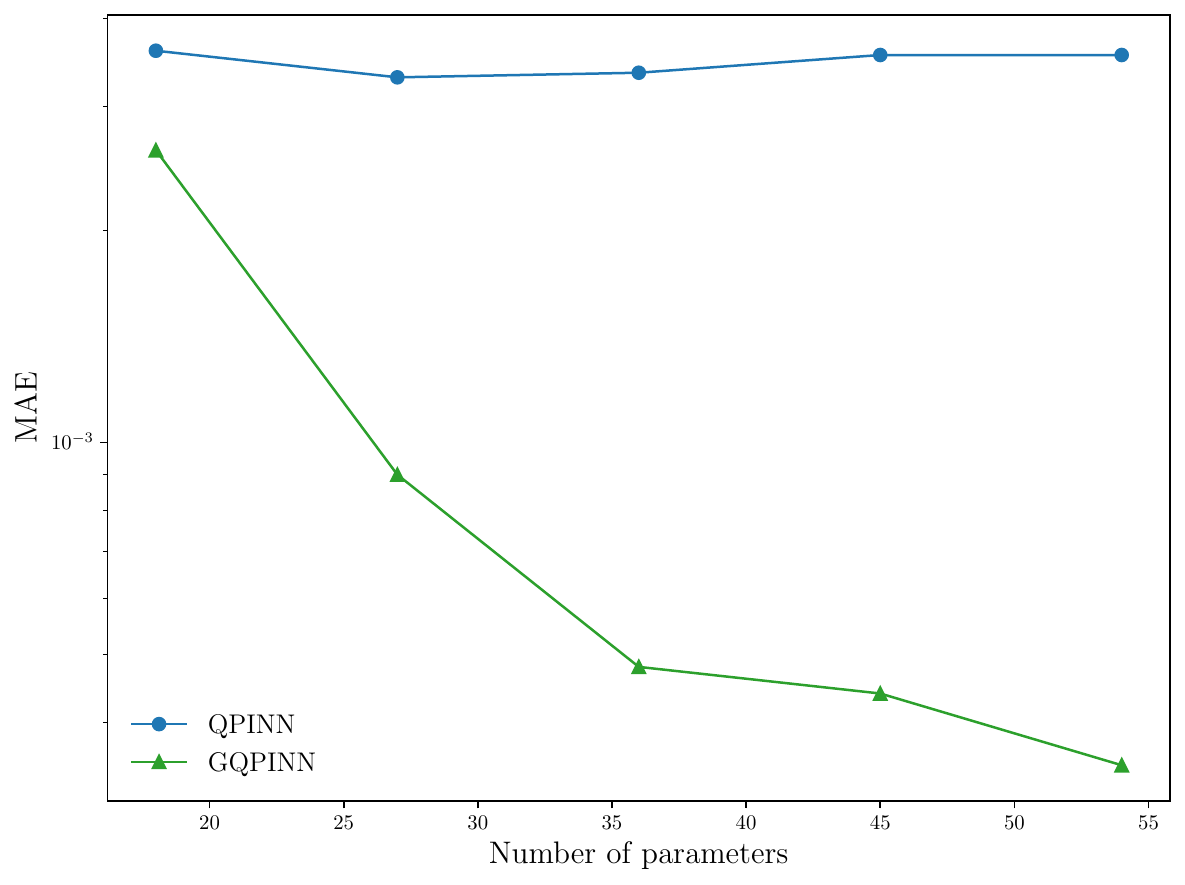} 
\caption{MAE of the QPINN baseline and the GQPINN ansatz (Fig.\ref{fig:diffusion_circuit}) on the 2D diffusion equation across different numbers of trainable parameters.}
\label{fig:benchmark_dynamic_depth}
\end{figure}

\begin{figure*}[t]
\centering
\includegraphics[width=\textwidth]{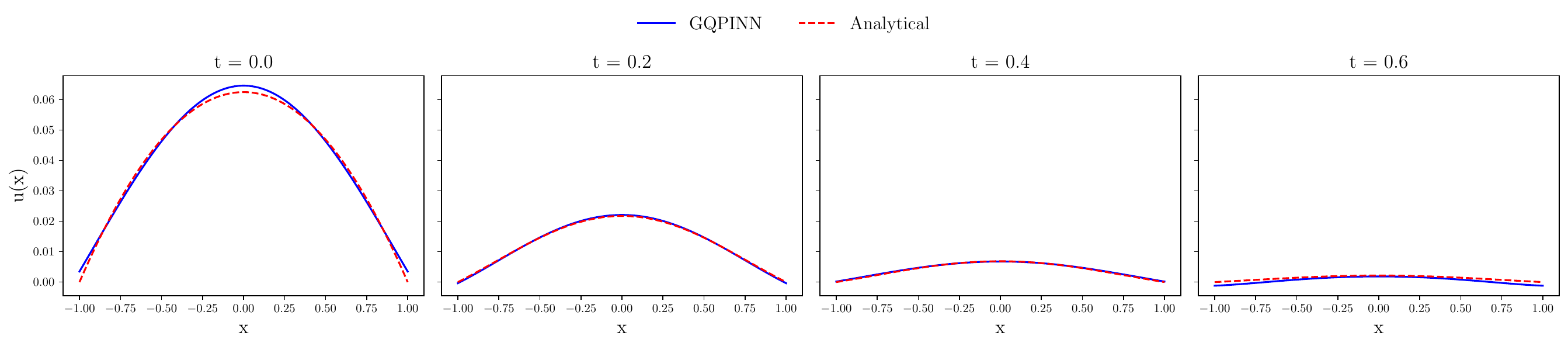} 
\caption{Predictions of the GQPINN ansatz (Fig.\ref{fig:diffusion_circuit}) (labeled GQPINN) compared with the analytical solution of the diffusion equation at $y=0$, using a 3-qubit, 5-layer circuit. Profiles are shown at $t=0$, $0.2$, and $0.4$ (training time domain) and at $t=0.6$ (an unseen time point). Results correspond to a representative run (one of 10 trials).}
\label{fig:pred_dynamic}
\end{figure*}

\begin{figure}[t]
\centering
\includegraphics[width=\linewidth]{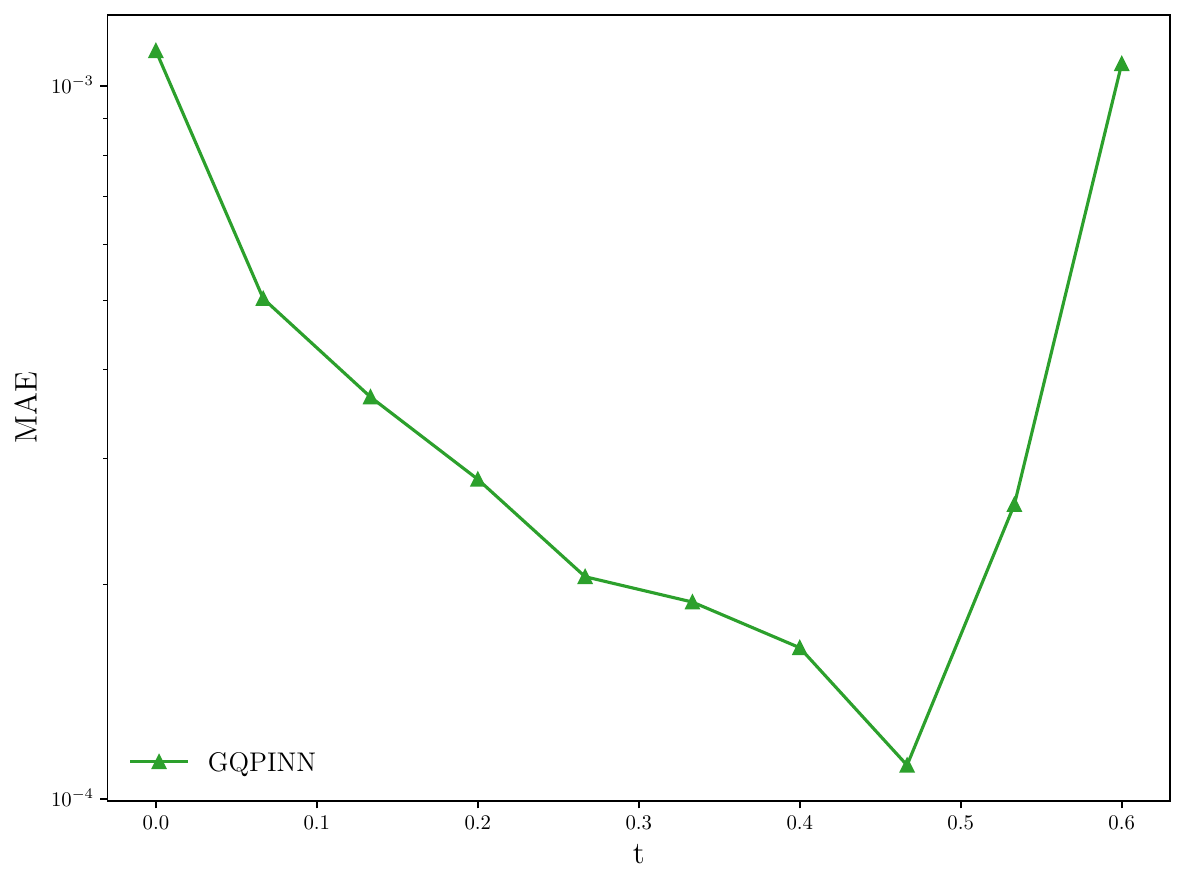}
\caption{MAE of the GQPINN ansatz (Fig.\ref{fig:diffusion_circuit}) prediction versus time $t$ for the diffusion equation using a 5-layer circuit. Results are shown as the average over 10 samples. The minimum error is attained for $t\in[0.4,0.5]$ (training time domain), while $t>0.5$ corresponds to unseen time points used to assess temporal generalization.}
\label{fig:mae_dynamic_time}
\end{figure}

Having established the advantage of continuous symmetry groups for the Poisson equation (the steady-state diffusion case), we now turn to the diffusion equation,
\begin{equation}
\frac{\partial^2 u}{\partial x^2} + \frac{\partial^2 u}{\partial y^2}
= \frac{1}{D}\frac{\partial u}{\partial t},
\label{diffusion}
\end{equation}
where the spatial operator remains rotational invariant about the $z$-axis. We further define $r=\sqrt{x^2+y^2}$ and consider a disk of radius $R$. Homogeneous Dirichlet boundary conditions are imposed by setting $u_{\text{exact}}(r,t)=0$ on $r=R$, and the initial condition is given by $u_{\text{exact}}(r,0)$. The analytical solution of this problem is

\begin{equation}
\begin{aligned}
u_{\text{exact}}(r,t)
&=
-\sum_{n=1}^{\infty}
\frac{R^{2} J_0\!\left(\alpha_{0n}\frac{r}{R}\right)}{2\,\alpha_{0n}^{3}\,J_{1}(\alpha_{0n})}\ e^{-\frac{D \alpha_{0n}^{2}}{R^{2}}t}.
\end{aligned}
\label{eq:uexact_bessel}
\end{equation}
Here $J_0(\cdot)$ and $J_1(\cdot)$ denote Bessel functions of the first kind of orders $0$ and $1$, respectively, and $\alpha_{0n}$ is the $n$th positive root of $J_0(\alpha)=0$. 
The disk radius and diffusion coefficient are set to $R=D=1$, and the series is truncated at $n=1,\ldots,50$. 

We employ $N_{\text{Res}}=1720$, $N_{\text{B}}=3800$, and $N_{\text{I}}=172$ training points to compute the total loss function.  The training points are selected as follows. For the residual points, we first sample $16$ evenly spaced points in $x\in[-1,1]$ and $16$ evenly spaced points in $y\in[-1,1]$, together with $10$ evenly spaced points in $t\in[0,0.5]$. Points outside the computational domain are then removed by filtering out locations satisfying $x^2+y^2>1$, resulting in $N_{\text{Res}}=1720$. For the boundary points, $200$ evenly spaced points are sampled along the circular boundary $x^2+y^2=1$, and $20$ evenly spaced points are sampled in $t\in[0,0.5]$, giving $N_{\text{B}}=3800$. For the initial-condition points, we directly reuse the spatial locations from the residual set $N_{\text{Res}}$ at the initial time $t=0$. This yields $N_{\text{I}}=172$ initial-condition points. We follow the same evaluation protocol as in the steady-state diffusion case. 
Specifically, we compare the MAE of the extension of $SO(2)$-GQPINN ansatz for time-dependent dynamics with that of the QPINN while fixing the number of qubits to 3 and increasing the number of layers from 1 to 5. The corresponding trainable blocks contain 9 parameters per block for this GQPINN model and 12 parameters per block for the QPINN baseline.

Fig.~\ref{fig:benchmark_dynamic_depth} compares the MAE of the GQPINN model with that of the QPINN ansatz. The GQPINN model performance improves consistently as the number of trainable parameters increases and begins to converge once the parameter count reaches the moderate regime, whereas the QPINN baseline largely stagnates even as the parameter count increases. This trend suggests that increasing model capacity is effective when symmetry is embedded into the ansatz, while simply adding parameters without symmetry does not provide the same benefit. This behavior is consistent with the case of Poisson equation, where the symmetry-embedded ansatz achieves lower MAE with fewer trainable parameters, indicating that the improvement is driven by the symmetry-induced inductive bias rather than model size.

Fig.~\ref{fig:pred_dynamic} shows the GQPINN model predictions from a 5-layer model for a representative run (one of the 10 training trials) compared with the analytical solution at the slice $y=0$ over $t=0$ to $t=0.6$. For the time domain used during training, $t\in[0,0.5]$, the predicted profiles closely coincide with the analytical curves, with particularly strong agreement at $t=0.4$. This trend is consistent with Fig.~\ref{fig:mae_dynamic_time}, where the MAE decreases over $t\in[0,0.4]$ and attains its lowest values in the interval $t\in[0.4,0.5]$, matching the visually tight overlap observed in the slice comparison. We further evaluate generalization by testing at $t=0.6$, which is not included in the training data. At this unseen time point, the model captures the expected diffusion behavior, producing a profile that continues to shrink in magnitude relative to earlier times. Although the discrepancy increases at $t=0.6$, the error remains on the order of $10^{-3}$, indicating that the model learns the underlying temporal dynamics reasonably well despite being trained only on earlier time snapshots.

\section{$\mathcal{G}$-equivariant models}
\label{app:Z_2_case}
In this section, we move beyond $\mathcal{G}$-invariant models and consider $\mathcal{G}$-equivariant constraints. 
As an example, we take $\mathcal{G}=\mathbb{Z}_2=\{e,p\}$, where $p$ acts on the spatial coordinate as $(x,t)\mapsto(-x,t)$, and impose odd parity under spatial inversion:
\begin{equation}
u(x,t) = -u(-x,t).
\label{eq:odd_parity}
\end{equation}
By choosing the nontrivial group element $g=p$ and $K_p=-1$, Eq.~\eqref{eq:equi_space_eq} for the $\mathbb{Z}_2$-equivariant model reduces to
\begin{equation}
    \expval{O}{\psi_{\bm{\theta}}(-x,t)}
    =-\expval{O}{\psi_{\bm{\theta}}(x,t)}.
\label{eq:z_2_equiv}
\end{equation}
Thus, Eq.~\eqref{eq:z_2_equiv} defines a $\mathcal{G}$-equivariant model with respect to the $\mathbb{Z}_2$ symmetry.

\begin{figure*}[t]
\centering
\begin{quantikz}
\lstick{$q_1$} & \qw  & \gate{H} \gategroup[2,steps=1,style={dashed,rounded
corners,fill=blue!20, inner
xsep=2pt},background,label style={label
position=below,anchor=north,yshift=-0.2cm}]{{\sc
initial state}} 
  & \qw
  & \gate[2]{R_{XX}(\theta_1)} 
  \gategroup[2,steps=3,style={dashed,rounded
    corners,fill=blue!20, inner
    xsep=2pt},background,label style={label
    position=below,anchor=north,yshift=-0.2cm}]{{\sc
    training block}}
  & \gate{R_X(\theta_2)} 
  &
  & \qw
  & \gate{R_Y(x)} 
  \gategroup[2,steps=1,style={dashed,rounded
    corners,fill=blue!20, inner
    xsep=2pt},background,label style={label
    position=below,anchor=north,yshift=-0.2cm}]{{\sc
    encoding block}}  
  & \qw
  \slice{} 
  & \push{\,\cdots\,} 
  & \qw
  \slice{} 
  & \qw
  & \gate[2]{W_{p+1}}
  \gategroup[2,steps=1,style={dashed,rounded
    corners,fill=blue!20, inner
    xsep=2pt},background,label style={label
    position=below,anchor=north,yshift=-0.2cm}]{{\sc
    training block}}  
  & \qw
  & \meter{} \\
\lstick{$q_2$} & \qw & \qw     
  & & & \gate{R_Y(\theta_3)}
  & \gate{R_Z(\theta_4)} &
  & \gate{R_Y(y)} &
  & \push{\,\cdots\,} 
  &  &  &  & 
  & \meter{}
\end{quantikz}
\caption{$\mathbb{Z}_2$-GQPINN ansatz enforcing equivariance under parity inversion $(x,t)\mapsto(-x,t)$. The inputs $(x,t)$ are encoded onto the qubits via angle encoding using $R_Y$ rotations. The trainable layer is constructed to be $\mathbb{Z}_2$-equivariant using the symmetry-preserving generator set in Eq.~\eqref{eq:z2_gates}, and an additional final trainable block is applied before measurement.}
\label{fig:equ_circuit}
\end{figure*}
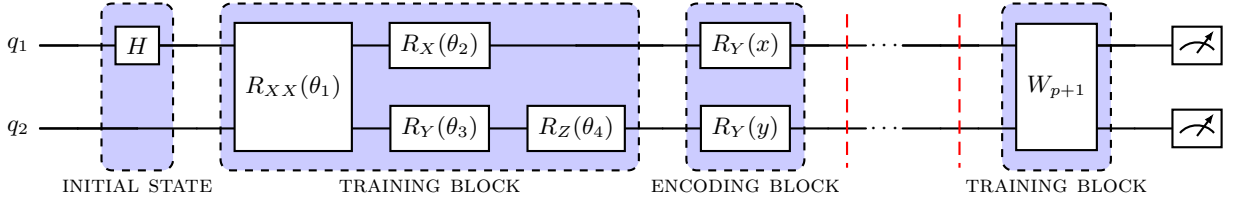

The construction of an $\mathcal{G}$-equivariant model is largely the same as that of a $\mathcal{G}$-invariant model, except that the observable must be chosen appropriately.
Using the same data-encoding unitary in Eq.~\eqref{eq:Ry_encoding}, $U(x,t)=R_Y(x)\otimes R_Y(t)$, the corresponding induced representations are given by $U_e = I\otimes I$ and $ U_p = X\otimes I$.
Following the same procedure as in Sec.~\ref{sec:discrete}, we obtain the commuting generator set for $\mathbb{Z}_2$ as
\begin{equation}
\mathcal{T}_{\mathbb{Z}_2}(\mathcal{S})
= \bigg\{ X_1,\ X_2,\ Y_2,\ Z_2,\ X_1X_2 \bigg\}.
\label{eq:z2_gates}
\end{equation}
We choose an invariant initial state given by a superposition on the first qubit, which encodes the classical spatial data, and the $\ket{0}$ state on the second qubit, i.e., $\ket{\psi_0}=\frac{1}{\sqrt{2}}(\ket{00}+\ket{10}).$ We further select the equivariant observable $O = (Z_1 + Y_1)Z_2$ so that it satisfies the odd-parity condition $U_p^\dagger O\, U_p = -O$. 
For simplicity, We refer to our $\mathbb{Z}_2$-equivariant GQPINN as $\mathbb{Z}_2$-GQPINN.
The complete $\mathbb{Z}_2$-GQPINN ansatz is illustrated in Fig.~\ref{fig:equ_circuit}. 

\begin{figure}[t]
\centering
\includegraphics[width=\linewidth]{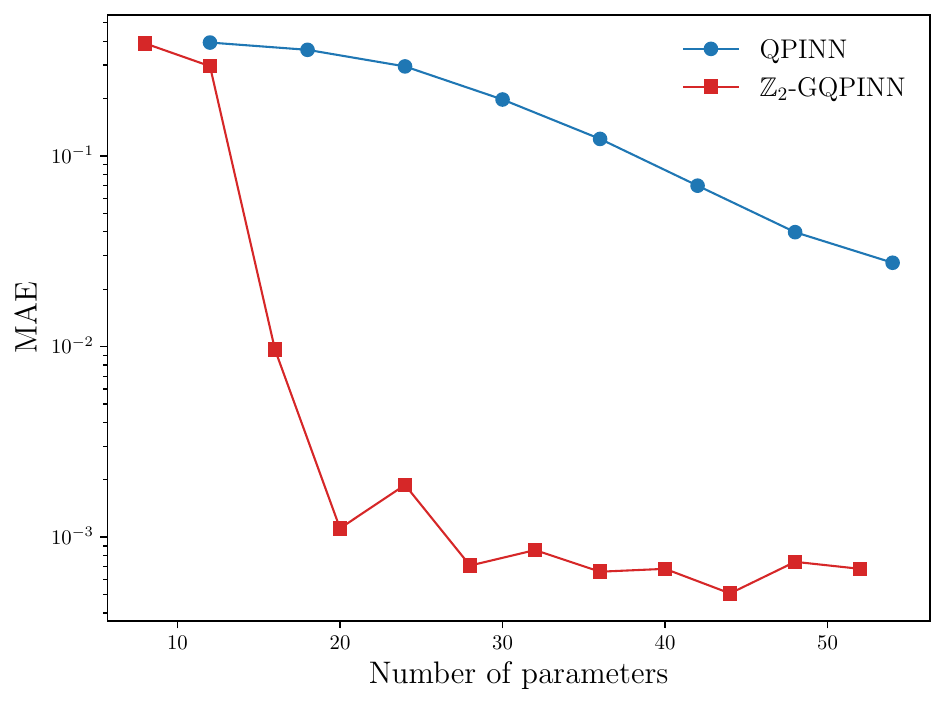}
\caption{MAE of the QPINN baseline and the $\mathbb{Z}_2$-GQPINN ansatz across different numbers of trainable parameters on the acoustic wave equation.}
\label{fig:mae_wave}
\end{figure}

\begin{figure}[t]
\centering
\includegraphics[width=\linewidth]{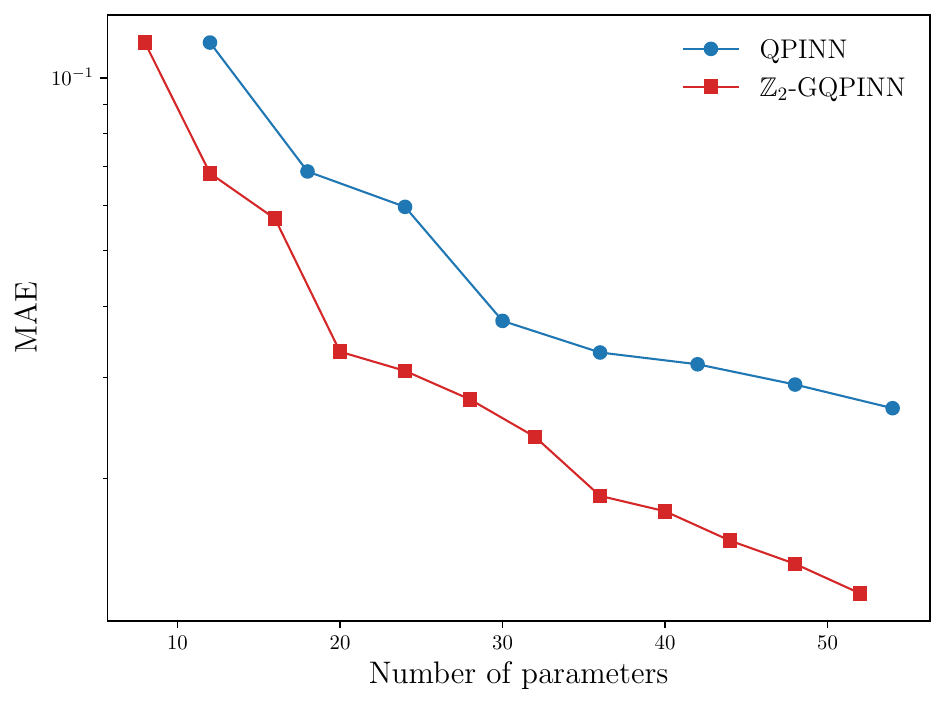}
\caption{MAE of the QPINN baseline and the $\mathbb{Z}_2$-GQPINN ansatz across different numbers of trainable parameters for the viscous Burgers' equation.}
\label{fig:mae_burger}
\end{figure}

Motivated by the success of GQPINN on the Poisson equation and the diffusion equation, which belong to the linear elliptic and parabolic PDE classes, we consider a linear hyperbolic model problem given by the one-dimensional acoustic wave equation 
\begin{equation}
    \frac{\partial^2 u}{\partial t^2} = c^2 \frac{\partial^2 u}{\partial x^2},
\label{eq:wave_pde}
\end{equation}
and a nonlinear parabolic PDE given by the viscous Burgers' equation
\begin{equation}
    \frac{\partial u}{\partial t} + u\frac{\partial u}{\partial x} = \nu \frac{\partial^2 u}{\partial x^2}.
\label{eq:burgers_pde}
\end{equation}
In both cases, we restrict to a one-dimensional spatial domain in this paper.
In this subsection, we investigate how the performance changes when $\mathbb{Z}_2$-equivariant GQPINNs are applied to these linear and nonlinear PDEs.

Since the 1D acoustic wave equation is second order in time, the loss function for Eq.~\eqref{eq:wave_pde} includes an additional initial-condition term that enforces the initial velocity, $\partial_t u_{\text{exact}}(x_i,t_0)$. The initial displacement and velocity at $t_0=0$ are specified from the analytical solution,
\begin{equation}
\begin{aligned}
u_{\text{exact}}(x,t) &= A \cos(kx)\cos(\omega t)\\
\frac{\partial u_{\text{exact}}(x,t)}{\partial t} &= -A\omega \cos(kx)\sin(\omega t),
\end{aligned}
\end{equation}
where $k=\omega/c$ is the wave number. Throughout the experiments, the speed of sound and the solution amplitude are fixed to $c=A=1$, and the wave number is set to $k=\pi$.

Meanwhile, the loss function for 1D Burgers' equation~\eqref{eq:burgers_pde} is defined in Eq~\eqref{QPINN_loss}. 
The initial condition at $t_0=0$ is set according to the analytic solution,
\begin{equation}
u_{\text{exact}}(x,t) =
\frac{\dfrac{x}{t+1}}
{1 + \sqrt{\dfrac{t+1}{t_0}}
\exp\!\left(\dfrac{x^2}{4\nu (t+1)}\right)},
\end{equation}
where $t_0 = \exp\!\left(\frac{1}{8\nu}\right)$ and $\nu$ is the viscosity. We set the viscosity $\nu=0.01$

For these two PDEs, we use the same sampling scheme. For the residual set, we sample $20$ evenly spaced points in $x\in[-1,1]$ and $10$ evenly spaced points in $t\in[0,1]$, yielding $N_{\text{Res}}=200$ collocation points. For the initial-condition set, we reuse the spatial locations from the residual grid at $t=0$, which yields $N_{\text{I}}=20$ initial-condition points. We then compare the MAE of the $\mathbb{Z}_2$-QPINN ansatz with that of the conventional QPINN ansatz while fixing the number of qubits to $2$ and increasing the number of layers from $1$ to $8$. The number of trainable parameters per layer is $4$ for the $\mathbb{Z}_2$-GQPINN ansatz and $6$ for the QPINN ansatz. All remaining training and evaluation settings follow the previous section.

Figure~\ref{fig:mae_wave} compares the MAE of the $\mathbb{Z}_2$-GQPINN ansatz with that of the conventional QPINN baseline for the 1D acoustic wave equation as a function of the number of trainable parameters. The $\mathbb{Z}_2$-GQPINN exhibits a markedly faster error reduction as the parameter counts increases, dropping by more than an order of magnitude once the parameter count enters the moderate regime around  $15-20$ parameters and then stabilizing around the $10^{-3}$ level with small fluctuations. In contrast, the QPINN baseline decreases much more slowly and remains in the $10^{-2}-10^{-1}$ range over the evaluated parameter ranges. Overall, these results indicate that embedding the $\mathbb{Z}_2$ equivariant constraint provides a strong inductive bias for the hyperbolic wave dynamics and substantially improves parameter efficiency.

Figure~\ref{fig:mae_burger} shows the analogous comparison for the 1D time-dependent viscous Burgers' equation. While the $\mathbb{Z}_2$-GQPINN consistently achieves lower MAE than the QPINN baseline across the tested range, the improvement is comparatively modest and both methods exhibit a gradual decrease in MAE as the number of parameters increases. Compared with the 1D wave equation and 2D Poisson equation benchmarks, the symmetry benefit is less pronounced for viscous Burgers. 
 This result suggests that learning nonlinear convection-diffusion dynamics is more difficult than linear partial differential equations.

\begin{figure}[]
\centering
\includegraphics[width=\linewidth]{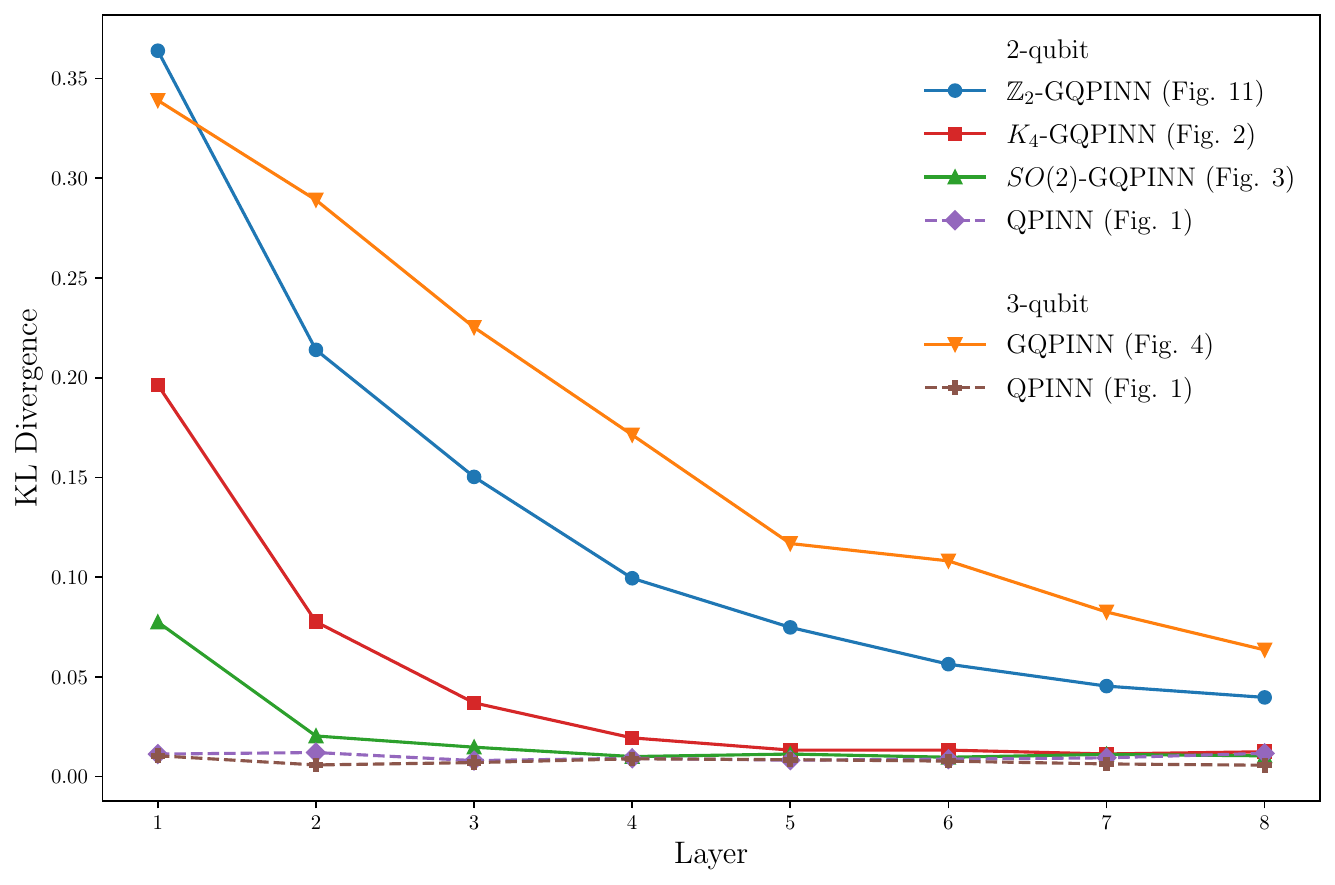} 
\caption{KL-divergence as a function of the number of layers $p$ for the quantum circuits shown in Figs.~\ref{fig:conv_pqc}, \ref{fig:sym_circuit}, \ref{fig:cont_sym_circuit}, \ref{fig:diffusion_circuit}, and \ref{fig:equ_circuit}.}
\label{fig:benchmark_KL}
\end{figure}

\section{Expressibility Measure}
\label{sec:expressibility}

A widely adopted approach for quantifying and comparing the expressibility of parametrized quantum circuits (PQCs) is to compute the Kullback--Leibler (KL) divergence~\cite{Sim_2019}:
\begin{equation}
    D_\mathrm{KL}\qty(P_\mathrm{PQC}^{\prime} (F;\Theta_M) ||P_\mathrm{Haar}(F)).
\end{equation}
Here, $\Theta_M$ denotes a pool of $M$ sampled circuit parameter configurations. 
In our QPINN setting, each element of $\Theta_M$ includes both the trainable parameters in the training blocks and the data inputs in the encoding block.
$F$ is the fidelity between two quantum states generated from configurations in this pool, and $P_\mathrm{PQC}^{\prime}(F;\Theta_M)$ denotes the empirical distribution of these fidelities. 
$P_{\mathrm{Haar}}(F)$ denotes the corresponding fidelity distribution for pairs of Haar-random pure states in the same Hilbert space. 
This metric evaluates how closely the fidelity distribution induced by a PQC matches that of Haar-random pure states.
In this work, we compute the KL divergence for various circuit architectures, both with and without imposed symmetries, illustrated in Figs.~\ref{fig:conv_pqc}, \ref{fig:sym_circuit}, \ref{fig:cont_sym_circuit}, \ref{fig:diffusion_circuit}, and \ref{fig:equ_circuit}, in order to compare their state-level expressibility.
To compute the KL divergence, we construct the empirical fidelity distribution from $5000$ sampled pairs of quantum states generated from the pool.
The trainable parameters are drawn uniformly from $[0,2\pi)$ and the data inputs are drawn uniformly from the problem-specific domains defined in Sections~\ref{sec:experimental_result} and~\ref{app:Z_2_case}.
We note that the invariance or equivariance of the model is not determined solely by the quantum circuit, since the observable must also be chosen to be compatible with the desired symmetry.
Therefore, the KL divergence of the state ensemble should be interpreted only as a state-level expressibility
measure, not as a direct certificate of output-level invariance or equivariance.

The results, shown in Fig.~\ref{fig:benchmark_KL}, reveal that both the 2- and 3-qubit QPINNs exhibit lower KL divergences.
A smaller KL divergence indicates greater expressibility with respect to the Haar distribution over the full Hilbert space.
However, this enhanced expressibility does not translate into superior performance on the target task.
These results suggest that expressibility measured via KL divergence relative to the full Hilbert space does not necessarily correlate with improved performance. 
The performance of a QPINN is inherently problem-dependent. 
In the PDE-solving tasks considered in this work, a quantum circuit need not be maximally expressive over the full Hilbert space.
Rather, the ansatz and observables should be chosen so that the model induces a function class suited to approximating solutions satisfying the target PDE and its initial and boundary conditions while respecting the imposed symmetries.
This highlights the importance of designing quantum circuits that explicitly incorporate the symmetries, structure, and constraints of the target problem~\cite{Safari2026Unpublished}.

\section{Summary And Outlook}
\label{sec:summary}

In this work, we propose geometric QPINN (GQPINN) ansatzes for solving PDEs and evaluate them on a range of representative linear and nonlinear problems, including the two-dimensional Poisson equation, the diffusion equation, the one-dimension acoustic wave equation, and the one-dimensional viscous Burgers’ equation. Spatial structure is incorporated by embedding symmetry constraints directly into the circuit architecture through equivariant generator sets derived from group representations via group averaging (twirling).

Across all benchmarks, the GQPINN ansatzes consistently outperform the baseline QPINN. For the 2D Poisson problem, both $K_4$-GQPINN and $SO(2)$-GQPINN achieve substantially lower MAE than QPINN, with $SO(2)$-GQPINN providing the best overall performance and reaching up to two orders of magnitude reduction in MAE while using fewer trainable parameters. When results in Fig.~\ref{fig:benchmark_static_param} are reported as a function of trainable parameters, performance improves rapidly in the low-to-moderate parameter count and typically begins to saturate once the parameter count reaches $\sim 20$, indicating diminishing returns from adding further parameters. Increasing the number of qubits from two to four does not yield a systematic advantage once the total number of trainable parameters is comparable, suggesting that the parameter count, rather than Hilbert-space dimension alone, is the dominant factor in this setting.
Classical comparisons further highlight the role of symmetry. Symmetry-invariant PINN (SI-PINN) improves upon a standard PINN and achieves low MAE at small parameter counts, although its performance does not improve further as the parameter count increases. Comparing SI-PINN and GQPINN models, the performance gap depends on the symmetry construction: $K_4$-GQPINN is competitive with the corresponding SI-PINN but does not provide a consistent improvement at comparable parameter counts, whereas $SO(2)$-GQPINN attains the lowest MAE overall. Taken together, these results suggest that symmetry constraints improve parameter efficiency in both classical and quantum models, and that the $SO(2)$-GQPINN ansatz achieves the best performance in this benchmark.

For the diffusion equation, the extension of $SO(2)$-GQPINN continues to improve as the number of trainable parameters increases, whereas the QPINN baseline shows little improvement, indicating that symmetry embedding remains effective in dynamic settings and acts as an inductive bias that reduces the search space during optimization. 

We further move beyond $\mathcal{G}$-invariant models and consider $\mathcal{G}$-equivariant constraints. For the 1D acoustic wave equation, the GQPINN achieves more than an order-of-magnitude reduction in MAE relative to QPINN at moderate parameter counts and stabilizes at a substantially lower error level. For the 1D viscous Burgers’ equation, GQPINN remains consistently better than QPINN, though the improvement is more modest, which is consistent with the increased difficulty of learning nonlinear convection–diffusion dynamics.

Taken together, these results demonstrate that symmetry embedding via twirling provides a practical and effective design principle for improving both trainability and parameter efficiency in quantum machine learning methods for solving PDEs. As future work, our current approach enforces symmetry primarily at the level of the target solution, for example rotational invariance. However, many PDEs admit richer point symmetries in the sense of Lie symmetry analysis~\cite{olver1993applications}, including translations and scalings. These symmetries act on the independent and dependent variables and generally map one solution of the PDE to another,
i.e., $u(x)\mapsto u'(x)$ with $u'(x)\neq u(x)$ in general, while both $u$ and $u'$ satisfy the same PDE. Incorporating such point symmetries in a unified way may require learning the solution operator (mapping inputs such as coefficients, sources, or initial/boundary conditions to the corresponding solution) rather than fitting a single solution instance. In classical deep learning, operator-learning frameworks such as DeepONet~\cite{Lu2021DeepONet} provide a natural starting point, and hybrid approaches that combine DeepONet with physics-informed losses~\cite{Wang2021DeepONet} can enforce PDE constraints without relying on labeled solution data. Extending symmetry-embedded QML to operator learning is therefore a promising direction for capturing broader PDE symmetries and improving generalization across problem instances.

\begin{acknowledgments}
We would like to thank Kuan-Cheng Chen for helpful comments on this research.
H.M. and W.T. acknowledge that part of this work was performed for Council for Science, Technology and Innovation (CSTI), Cross-ministerial Strategic Innovation Promotion Program (SIP), “Promoting the application of advanced quantum technology platforms to social issues” (Funding agency : QST). R.S. thanks DLR Quantum Computing Initiative and the Federal Ministry for Economic Affairs and Climate Action; qci.dlr.de/projects/toquaflics.
\end{acknowledgments}

\appendix
\section{Symmetry-Invariant Neural Network}
\label{app:sinn}
To construct a classical symmetry-aware counterpart for comparison, symmetry must be incorporated into a PINN in a way that is as comparable as possible to our circuit-level construction. Broadly, there are two common strategies. The first is a loss-based approach~\cite{Zhang2022SPINN}, where symmetry constraints are imposed as additional penalty terms in the PINN loss function. The second is an architecture-based approach~\cite{Zhu20221SPINN,Zhang2025sDNN}, where the network is modified so that invariance/equivariance is satisfied by construction. Because GQPINN enforces symmetry through the circuit architecture (rather than introducing an additional symmetry-specific loss term), the architecture-based strategy provides the closest classical analogue in our setting.

Inspired by the a symmetry-enhanced deep neural network (sDNN) in~\cite{Zhang2025sDNN}, we propose a new design that enforces invariance only at the network input by averaging the first-layer features computed from each  group-transformed input $V_g [\bm{x}]$ ($g \in \mathcal{G}$). After this group-averaged input embedding step, the architecture follows a standard multilayer perceptron (MLP), and the final output is a single scalar. In contrast to the GQPINN construction, this SI-PINN design does not reduce the number of trainable parameters relative to a standard PINN with the same number of hidden nodes. 

Let $W_1\in\mathbb{R}^{m\times n}$, $\mathbf{b}_1\in\mathbb{R}^m$, denote trainable parameters, and let $\phi(\cdot)$ denote the activation function applied element-wise, where $m$ is the number of hidden units and $n$ denotes the input dimension. The group-averaged input embedding (first hidden layer) is defined as
\begin{equation}
\mathbf{h}^{(1)}(\mathbf{x})
=
\frac{1}{|\mathcal{G}|}
\sum_{g\in \mathcal{G}}
\phi\!\left(\mathbf{W}_1\,(V_g[\bm{x}]) + \mathbf{b}_1\right).
\label{eq:inv_lift}
\end{equation}
After the first layer, the network follow a standard MLP architecture with $\phi(\cdot)=\tanh(\cdot)$ activations, and the network produces a scalar output through a final linear readout. Although the forward pass evaluates group-transformed inputs $V_g[\bm{x}]$ $(g\in \mathcal{G})$, these branches share the same trainable weights $(\mathbf{W}_1,\mathbf{b}_1)$ and are combined through a group-averaging step, so the parameter count remains unchanged.

\bibliographystyle{apsrev4-2}
\bibliography{references}

\end{document}